\documentclass[aps,pre,showpacs,amsmath,amssymb,amsfonts,lengthcheck,twocolumn,longbibliography,superscriptaddress]{revtex4-2}

\usepackage{changes} 
\usepackage{ulem} 

\usepackage{graphicx}
 \usepackage{subfigure}
\usepackage{amsthm}
\usepackage{mathtools}
  \usepackage[utf8]{inputenc}
\usepackage{amsmath}
\usepackage{verbatim}
\usepackage{dcolumn}
\usepackage{bm}
\usepackage{color}
\usepackage[colorlinks=true,citecolor=blue,linkcolor=blue,urlcolor=blue]{hyperref}%
   \usepackage{xcolor}
\usepackage{physics}
\usepackage{dsfont}
 \usepackage{todonotes}
 
\usepackage{longtable}
 \usepackage{xargs}                      

\newcommand{\rescale}[2][1]{\resizebox{#1\textwidth}{!}{$#2$}}
\NewDocumentCommand\eq{ms}{%
  \IfBooleanTF{#2}{\begin{align}#1\end{align}}{\begin{align}
#1
\end{align}}
}
 
  \newcolumntype{M}[1]{>{\centering\arraybackslash}m{#1}}

\definecolor{rojo}{RGB}{255, 0, 0}

 \graphicspath{ {images/} }

\begin{document}

\title{Effects of Magnetic Anisotropy on 3-Qubit Antiferromagnetic Thermal Machines.}

\author{Bastian Castorene}
\email{bastian.castorene.c@mail.pucv.cl}
\affiliation{Instituto de Física, Pontificia Universidad Católica de Valparaíso, Casilla 4950, 2373223 Valparaíso,
Chile}
\affiliation{Departamento de Física, Universidad Técnica Federico Santa María, 2390123 Valparaíso, Chile}
\author{Francisco J. Pe\~na} 
\affiliation{Departamento de Física, Universidad Técnica Federico Santa María, 2390123 Valparaíso, Chile}
\affiliation{ Millennium Nucleus in NanoBioPhysics (NNBP), Av. Espa\~na 1680, Valpara\'iso 2390123, Chile}

\author{Ariel Norambuena} 
\affiliation{Centro Multidisciplinario de Física, Universidad Mayor, Camino la Pirámide 5750, Huechuraba Santiago, Chile}

\author{Sergio E. Ulloa} 
 \affiliation{Department of Physics and Astronomy and Nanoscale and Quantum Phenomena Institute, Ohio University, Athens, Ohio 45701, USA}
\author{Cristobal Araya} 
\affiliation{Departamento de Física, Universidad Técnica Federico Santa María, 2390123 Valparaíso, Chile}
 \author{Patricio Vargas}
\affiliation{Departamento de F\'isica, CEDENNA, Universidad T\'ecnica Federico Santa Mar\'ia,  Av. Espa\~na 1680, Valpara\'iso 2390123, Chile}

\date{\today}

\begin{abstract}
This study investigates the anisotropic effects on a system of three qubits with chain and ring topology, described by the antiferromagnetic Heisenberg XXX model subjected to a homogeneous magnetic field. We explore the Stirling and Otto cycles 
and find that {easy-axis} anisotropy significantly enhances engine efficiency across all cases. At low temperatures, the ring configuration outperforms the chain on both work and efficiency during the Stirling cycle. Additionally, in both topologies, the Stirling cycle achieves Carnot efficiency with finite work at quantum critical points. In contrast, the quasistatic Otto engine {also} reaches Carnot efficiency at these points but yields no useful work. Notably, the Stirling cycle exhibits all thermal operational regimes—engine, refrigerator, heater, and accelerator—unlike the quasistatic Otto cycle, which functions only as an engine or refrigerator.

\end{abstract}

\maketitle
\section{Introduction}\label{intro}
Quantum heat engines (QHEs) have become the focus of intense research due to their potential for developing highly efficient nanoscale devices operating with quantum working substances. These devices are characterized by specific thermodynamic cycles and operation dynamics. \cite{WOS:A1992HE74300067, WOS:A1992JN14600053, WOS:000174548900068,WOS:000247624300019,WOS:000246890100029,WOS:000170647500068,WOS:000259449100016,WOS:000460663800003,WOS:000238758700035,WOS:000339078100001, WOS:000228752500014,WOS:000251858800016,WOS:000243928200018,WOS:000178091500025, WOS:000577236900006,WOS:000360065300001,WOS:000074107900005,WOS:000359128100004,WOS:000280233300002, WOS:000377262900029, WOS:000341247200002,PhysRevA.62.062314}. The Stirling and Otto cycles are frequently encountered among the various thermodynamic machines \cite{ WOS:000400579500001,WOS:000592899500001,WOS:000451308800067, WOS:000544256400007, WOS:000426262900001,WOS:000446164500002,WOS:000506846500003,WOS:000630746700001, WOS:001054373800001,WOS:000452093900005,pena1}. When an external magnetic field is the driving parameter in quantum thermodynamics, the Stirling cycle consists of two isothermal and two isomagnetic strokes. During the isothermal strokes, the working substance interacts with thermal reservoirs at different temperatures as the magnetic field is varied, while in the isomagnetic strokes, the magnetic field remains constant.  In contrast, the Otto cycle comprises two isomagnetic and two adiabatic processes.  {Throughout} the isomagnetic quasistatic processes, the system interacts with different thermal reservoirs, whereas the magnetic field undergoes variation during the adiabatic strokes.  The efficiency of these cycles is notably influenced by the intensity of the external magnetic field, as it directly impacts the energy spectrum and states of the working substance. Consequently, the Stirling and Otto cycles in quantum thermodynamics driven by external magnetic fields exhibit promising potential for applications in magnetic refrigeration and related fields \cite{WOS:000832774500001, WOS:000425270000003, WOS:000778793500034}.
A recent investigation by Kuznetsova et al.\ \cite{Kuznetsova} studies a two-qubit Heisenberg XYZ model under a nonuniform magnetic field as the working medium in an Otto thermal machine.  The model incorporates Dzyaloshinskii–Moriya (DM) and Kaplan–Shekhtman–Entin-Wohlman–Aharony (KSEA) interactions, which are found to shift
the system's behavior as the longitudinal exchange constant transitions from antiferromagnetic to ferromagnetic values.
Distinct operational modes of the thermal machine were also identified,
including as heat engine, refrigerator, heater or dissipator, and thermal accelerator or cold-bath heater regimes.
Another study \cite{KamtaStarace} explores the impact of magnetic fields and anisotropy in exchange constants within a two-qubit XY model. The synergistic effect of the exchange anisotropy and magnetic field induces entanglement in parameter space regions previously devoid of it under isotropic conditions. This highlights the potential for entanglement control and generation in two-spin systems, even at finite temperatures.
Recently, a quantum Stirling cycle based on two coupled spins near a quantum critical point (QCP) has been investigated employing the XX isotropic Heisenberg model with a magnetic field \cite{WOS:000832774500001}. The study delves into the system's quantum phase transition, entanglement, and correlations, demonstrating that tailored cycle parameters enable the system to function as a heat engine or refrigerator across a broad range of magnetic field and low-temperatures. While the Stirling heat engine approaches Carnot efficiency near the critical point under high magnetic fields, the refrigerator cycle nears the Carnot limit for low magnetic fields. Nevertheless, performance deviates notably from Carnot limits at elevated temperatures, and maximum work output does not align with maximum efficiency.
Our recent work \cite{WOS:001130894000001} has shown that introducing magnetic anisotropy along the y-direction in the same model adjusts the maximum efficiency (Carnot) points in the cycle control parameters towards either lower or higher magnetic fields, contingent on the sign of the anisotropy. Therefore, including magnetic anisotropy improves machine performance from a thermodynamic point of view. 
 {Recent studies have proposed measuring tripartite entanglement in triangular systems \cite{WOS:000677559400001}. Additionally, three-spin qubit systems, such as those based on silicon, have been manufactured, and their stability and fidelity have been demonstrated experimentally \cite{WOS:000658600600002}. Investigations of a three-qubit quantum heat engine have been conducted using a Heisenberg XX model \cite{HeHeZheng}. It was found that while the efficiency approaches Carnot limits, it falls short of this value when the external field decreases.
 {Moreover, environment-induced quantum synchronization has been analyzed for a three-qubit configuration with different topologies~\cite{PhysRevA.101.042121}}.
}
    In this study, leveraging the effect of anisotropy to extend entanglement regions, we present an analysis of Stirling and Otto cycles employing a Heisenberg XXX description for a three-qubit system subjected to an external magnetic field and influenced by uniaxial magnetic anisotropy along the total spin in the y-direction.  For convenience, our description uses the inter-qubit exchange constant as the energy unit  {even though }
     experiments have determined the values of the exchange constant to be on the order of a few $\mu$eV in quantum dot qubits in silicon and two-hole spin qubits in thin field-effect transistors \cite{WOS:000501493200001, datos_2005, datos_2021, nature_ariel}. 
       Our investigation here focuses on evaluating the product of work and efficiency in magnetic Stirling and Otto cycles, specifically considering the role of magnetic anisotropy. We explore how anisotropy influences the temperature range where efficiency peaks, the magnetic field intervals that exhibit high efficiency, and the overall impact of anisotropy on these machines. We find that negative (easy-axis) anisotropy results in higher efficiency in both cycles. In contrast, positive anisotropy results in Stirling cycles reaching all four operational regimes, while only two in Otto cycles. We also find the ring topology to outperform the open chain configuration. Notably, Carnot efficiency is accompanied by finite work in the Stirling cycle for a ring system, whereas in the Otto cycle, it occurs with no work output.
    {For the Stirling cycle, achieving Carnot efficiency with finite work results from the conversion of the two isothermal strokes into adiabatic strokes}, {thereby transitioning the cycle into a Carnot cycle. This transformation occurs exclusively {at} the quantum critical point and under the selected parameters. This underscores the robustness of classical thermodynamics when applied to substances with quantum mechanical properties. } 
 \section{Model}
The working medium consists of three 
{qubits} described by a Heisenberg XXX {model} with nearest-neighbor interactions, uniaxial magnetic anisotropy along the y direction, {and a magnetic field along the z direction}
{
\begin{equation}\label{H}
    \mathcal{H} = J\qty(\sum_{i=1}^2 \boldsymbol{\hat{\sigma}}_i \cdot \boldsymbol{\hat{\sigma}}_{i+1}+ \alpha \qty(\boldsymbol{\hat{\sigma}}_1\cdot \boldsymbol{\hat{\sigma}}_3))+B \sum_{i=1}^3 \hat{\sigma}_i^z+   K S_y^2,
\end{equation}}
where the Pauli matrices \(\boldsymbol{\hat{\sigma}}_i = (\hat{\sigma}_i^x, \hat{\sigma}_i^y , \hat{\sigma}_i^z )\) represent the spin at the \(i_{\text{th}}\) site,
{and $S_y=\hat{\sigma}_1^y +\hat{\sigma}_2^y+\hat{\sigma}_3^y$ is the total Pauli operator along the anisotropy axis}. The exchange coupling constant $J$ will be used as unit of energy, and set to 1 ($-1$) for an antiferromagnetic (ferromagnetic) configuration. The {binary variable $\alpha$} takes the value of one (zero) for {the ring (chain) topology}.
\(B\) represents the external magnetic field along the \(z\) axis, and \(K\) is the anisotropy term. When \( K <0 \), we observe easy-axis anisotropy, which maximizes magnetization along the y-axis, while for \( K >0 \), one has easy-plane anisotropy, maximizing magnetization on the x-z plane.
\begin{figure*} 
\centering
\includegraphics[width=0.7\textwidth]{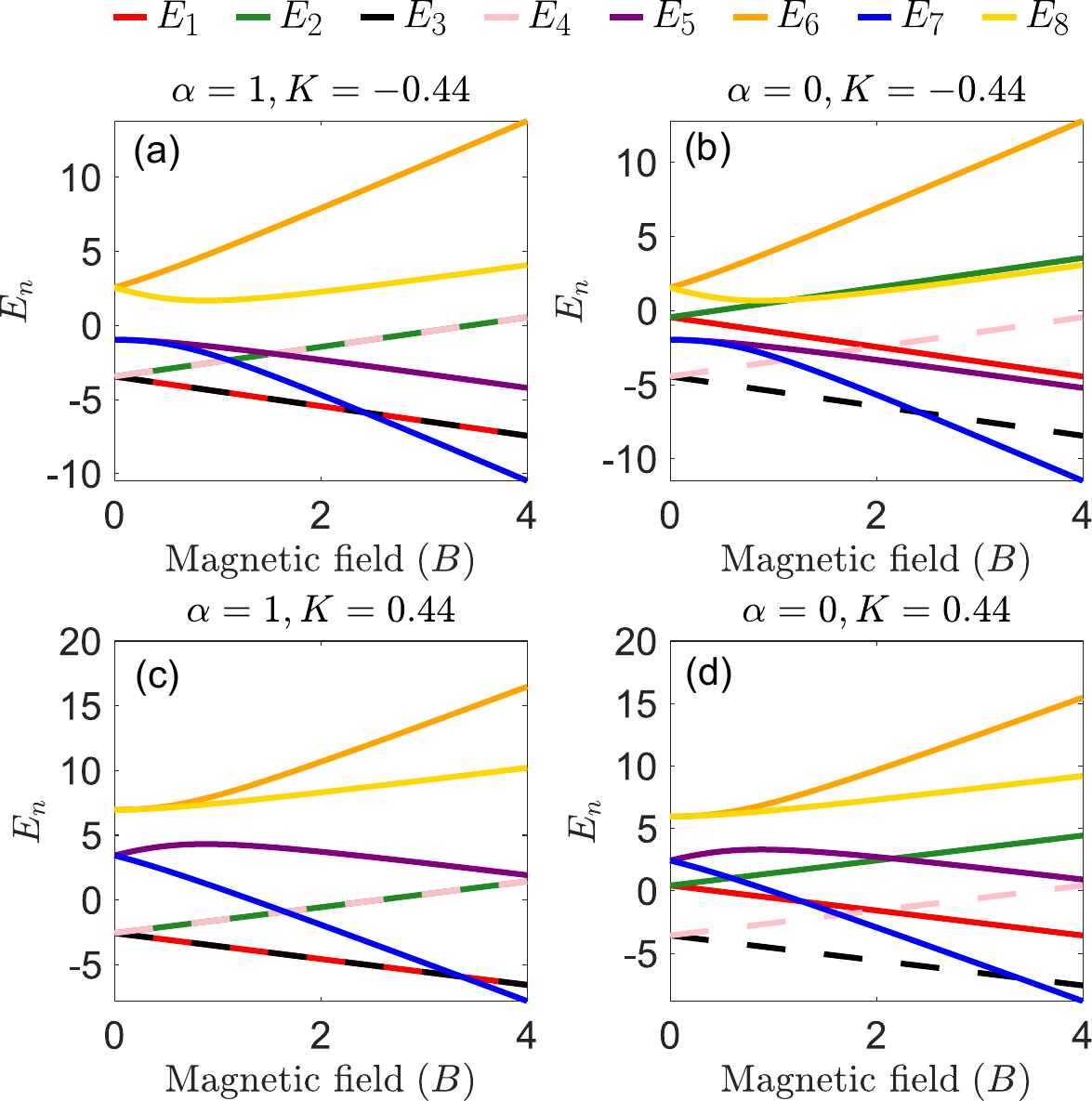}
\caption{{Energy spectrum of the antiferromagnetic system for different anisotropy $K$ and topologies ($\alpha=0,1$) as function of the magnetic field $B$. In (a) and (b), we observe the effect of negative anisotropy. The spectra shown in (c) and (d) are for positive anisotropy $K$; notice the quantum critical point is shifted to a higher field. The degeneracy ($E_1 = E_3$ and $E_2=E_4$) is only present in the ring topology; see (a) and (c). In these numerical calculations, we use $J=1$.}}
\label{PlotdeEnergias}
\end{figure*} 

The eigenvalues $E_n$ for the Hamiltonian  are given as:
\eq{
\begin{aligned}
E_1 &= -B - 3 \alpha J  +K, \\
E_2 &=  B - 3 \alpha J  +K, \\
E_3 &= {E_1 +4(\alpha-1)J}, \\
E_4 &= {E_2 +4(\alpha-1)J}, \\
E_5 &= B + J_{\text{eff}} - 2P_-,\\
E_6 &= B + J_{\text{eff}} + 2P_-,\\  
E_7 &=  -B +J_{\text{eff}}-2 P_+,  \\  
E_8 &= -B+J_{\text{eff}}+2P_+,
\end{aligned} \label{eigenenergies}
}
{where $P_\pm = \sqrt{B^2\pm2BK+4K^2}$ and $J_{\text{eff}} = (2+\alpha)J+5K$. A simple consequence of the topology of the system on the spectrum is that for $\alpha = 1$ (ring arrangement), the degeneracy $E_1 = E_3$ and $E_2 = E_4$ is present, contrary to the chain ($\alpha = 0$)}. The corresponding eigenvectors in the three-qubit computational $\sigma_z$ basis representation are: 
 {\eq{
\begin{aligned}
\ket{\psi_1} & =   \frac{1}{\sqrt{2}}\qty(  \ket{110}-\ket{011})\\
\ket{\psi_2} & =   \frac{1}{\sqrt{2}}\qty(\ket{100} - \ket{001})\\
\ket{\psi_3} & =   \frac{1}{\sqrt{6}} \qty(\ket{110}+\ket{011}-2 \ket{101})\\
\ket{\psi_4} & =  \frac{1}{\sqrt{6}} \qty(\ket{100}+\ket{001}-2 \ket{010}) \\
\ket{\psi_5}&= L_0^-\qty( \frac{R_0^-}{ K} \ket{000} + \ket{011}+ \ket{101}+\ket{110} )\\
{\ket{\psi_6}} & {= \frac{L_0^{-}}{\sqrt{3}}\left(\frac{R_0^{-}}{K}(|011\rangle+|101\rangle+|110\rangle)-3|000\rangle\right)} \\
\ket{\psi_7} &=  \frac{L_1^+}{\sqrt{3}} \qty(3\ket{111} - \frac{ R_1^+}{K} \qty(\ket{100}+\ket{010}+\ket{001}))     \\
\ket{\psi_8}&= \frac{L_0^+}{\sqrt{3}} \qty(3\ket{111} - \frac{ R_0^+}{K} \qty(\ket{100}+\ket{010}+\ket{001}))   .
\end{aligned}
}}
The various auxiliary variables are detailed in the appendix, Sec. \ref{variable_change}. {We note that { $\ket{\psi_1} = \ket{\Psi_{13}^-}\otimes \ket{1_2}$ and $\ket{\psi_2} = \ket{\Psi_{13}^-}\otimes \ket{0_2}$, where $\ket{\Psi_{13}^-} = (\ket{1_1,0_3}-\ket{0_1,1_3} )/\sqrt{2}$} is a Bell state for the bipartite system composed of the qubits 1 and 3.
$\ket{\psi_1}$ and $\ket{\psi_2}$ states are thus not fully tripartite entangled states (such as GHZ and $W$ states) but belong to the class of bipartite entanglement \cite{PhysRevA.62.062314}. Notice also that states $\ket{\psi_i}$ ($i=1,2,3,4$) have constant coefficients and are then robust against changes in magnetic field, exchange interaction, anisotropy, and topology. On the contrary, states \(\ket{\psi_5}\) to \(\ket{\psi_8}\) are influenced by changes in all physical parameters}. {The field dependence of the energy levels} results in a crossing point \(E_3 = E_7\) (\(E_1 = E_3 = E_7\)) for the antiferromagnetic chain (ring) at a critical field. Such level crossings mark quantum critical points (QCPs), where the ground state changes its dominant total $S_z$ character from $+1/2$ to $-1/2$, {which becomes relevant in the context of quantum heat engines \cite{WOS:000832774500001,WOS:001130894000001}, and will be crucial in order to maximize thermodynamical efficiency.}

The states at zero magnetic field exhibit the expected double degeneracy due to time-reversal symmetry {($\hat{\sigma}_i \rightarrow -\hat{\sigma}_i$)} for both chain and ring configurations. While the levels disperse with {the} magnetic field, the QCP can be obtained analytically, occurring at the critical field \(B_\text{crit} = -K + \sqrt{9J^2 + 12JK + K^2}\), independent of the system's topology. The main difference is the degenerate transition mentioned previously. 

To illustrate the impact of {the} anisotropy {and topology} on the system, {the} energy levels are depicted in Fig.~\ref{PlotdeEnergias} vs $B$ field, for anisotropy values of $K=-0.4$ and $K=0.4$. The anisotropy induces a shift in the quantum critical field to smaller values for negative \(K\) and to larger field for positive $K$. Notice also that negative \(K\) values narrow the gaps between energy levels, whereas positive \(K\) values expand them. These changes in level spacing due to anisotropy significantly influence the level population during magnetic cycles, which in turn modifies the operation of the heat machines.  This phenomenon will be further explored in subsequent sections.

The calculation of magnetic cycles consider that 
the thermodynamic quantities are fully defined by the canonical partition function of the system,  
$Z  = \sum_{i=1}^8 \exp \left(-E_i / T\right)$, {where we use the Boltzmann constant $k_B=1$ as a convention}.
The internal energy \(U\), Helmholtz free energy \(F\), and entropy \(S\) can be determined through the standard thermodynamic relations on the canonical ensemble:
\eq{U=T^2 \frac{\partial \ln Z}{\partial T}  , \quad F=-T \ln Z , \quad S=-\frac{\partial F}{\partial T } .}

 \section{Magnetic heat machines}
\subsection{Quantum Stirling Cycle}
 \begin{figure} 
\centering
\includegraphics[width=.45\textwidth]{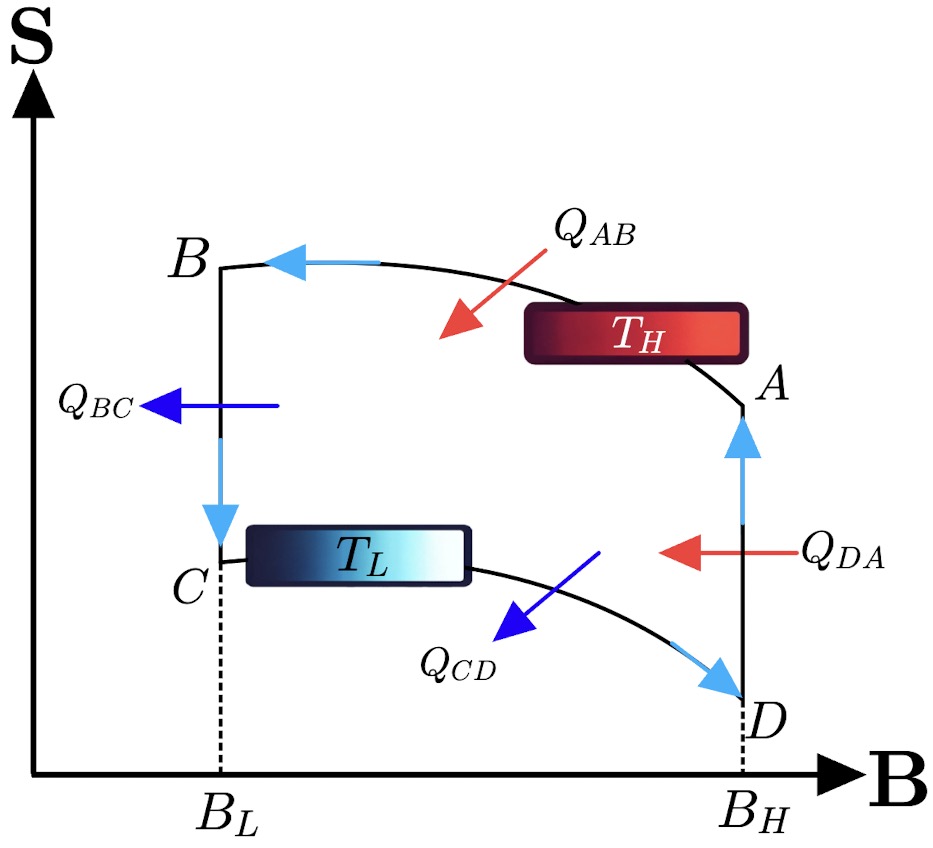}
\vspace{-4mm}
\caption{ \small Diagram of the Stirling cycle regarding entropy and external magnetic field along the z-axis. Curve \text{AB} is an isotherm at $T=T_H$ and curve \textit{CD} at $T=T_L$. }
\label{ciclo_stirling}
\end{figure}

 Mirroring its classical counterpart, the quantum Stirling cycle comprises four stages: two isothermal and two isomagnetic processes. The isomagnetic stages are analogous to the isochoric stages in the classical gas Stirling cycle.
 Throughout the isothermal processes, the system maintains thermal equilibrium with one of two thermal reservoirs, characterized by temperatures \(T = T_H\) and \(T = T_L\), where \(T_H > T_L\). Concurrently, the external magnetic field changes between \(B_H\) and \(B_L\), with \(B_L < B_H\). In the isomagnetic stages, the system undergoes a process at constant magnetic field intensity (either at \(B_L\) or \(B_H\)) where no work is performed. This cycle is illustrated in Fig.~\ref{ciclo_stirling}, with the external magnetic field $B_z$ as the control parameter. We analyze the cycle stage-by-stage, as presented in the figure. Note the system parameters, $J$, $\alpha$ and $K$ remain fixed throughout.
\begin{enumerate}
  \item[I.] Isothermal processes: These stages are represented by trajectories \(A \rightarrow B\) and \(C \rightarrow D\) in Fig.~\ref{ciclo_stirling}, during which the system is maintained at a constant temperature \(T_H\), \(T_L\), respectively, with entropy variations due to the energy exchange with the reservoir. The heat exchanged by the system can be ascertained from the entropy variation during the processes as:
\eq{ 
  Q_{A B}&=T_H\left(S_B(T_H,B_L)-S_A(T_H,B_H)\right),\\
Q_{C D}&=T_L\left(S_D(T_L,B_H)-S_C(T_L,B_L)\right).  }
\item[II.] Isomagnetic processes: These stages correspond to the trajectories \(B \rightarrow C\) (with \(B_L = \text{constant}\)) and \(D \rightarrow A\) (with \(B_H = \text{constant}\)) depicted in Fig.~\ref{ciclo_stirling} Throughout these processes, no work is performed; hence, the heat exchange is derived from the change in the internal energy as the system transitions between thermal baths from a temperature of \(T_H\) to \(T_L\) in the \(B \rightarrow C\) stage, and reversely from \(T_L\) to \(T_H\) in the \(D \rightarrow A\) stage. The heat exchanges during these stages are then:
\eq{
Q_{B C} & =U_C(T_L,B_L)-U_B(T_H,B_L), \\
Q_{D A} & =U_A(T_H,B_H)-U_D(T_L,B_H) . }
\end{enumerate}
From the first law of thermodynamics, we know that the variation of internal energy in a closed cycle is zero. Consequently, the total work is given by the expression:
\eq{W=Q_H+Q_L=Q_{A B}+Q_{D A}+Q_{B C}+Q_{C D} \, ,}

{where $Q_H=Q_{A B}+Q_{D A} > 0$ and $Q_L= Q_{C D}+Q_{B C} < 0$} is the heat received from and released to the reservoirs.
We adhere to the following sign convention: a positive (negative) value of \(Q\) denotes the system's absorption (release) of heat, while a positive (negative) value of \(W\) signifies work performed by (on) the system. Consequently, in the context of a heat engine, the conditions \(Q_H > 0\), \(Q_L < 0\), and \(W > 0\) prevail, as the system draws heat \(Q_H\) from the hot reservoir, releases a portion \(Q_L\) to the cold reservoir, and utilizes the remainder to execute an amount of work \(W\). The efficiency \(\eta\) is then defined as the ratio of the total work output to the heat absorbed by the system, expressed as:

{
\begin{eqnarray}
\eta=\frac{W}{Q_H}=1-\left|\frac{Q_{C D}+Q_{B C}}{Q_{A B}+Q_{D A}} \right|,
\end{eqnarray}}

 \subsection{Quasistatic Otto Cycle}
The cycle is thermodynamically defined by heat reservoirs and two distinct values of the external field, \(B_h\) and \(B_l\). Typically, the stage with lowest temperature \(\left(T_l\right)\) is associated with the lower field value \(\left(B_l\right)\), while the stage with highest temperature \(\left(T_h\right)\) corresponds to the higher field value \(\left(B_h\right)\). The thermodynamic processes that describe this cycle in Fig.~\ref{ciclo_otto} are the following: 
  \begin{figure} 
\centering
\includegraphics[width=.45\textwidth]{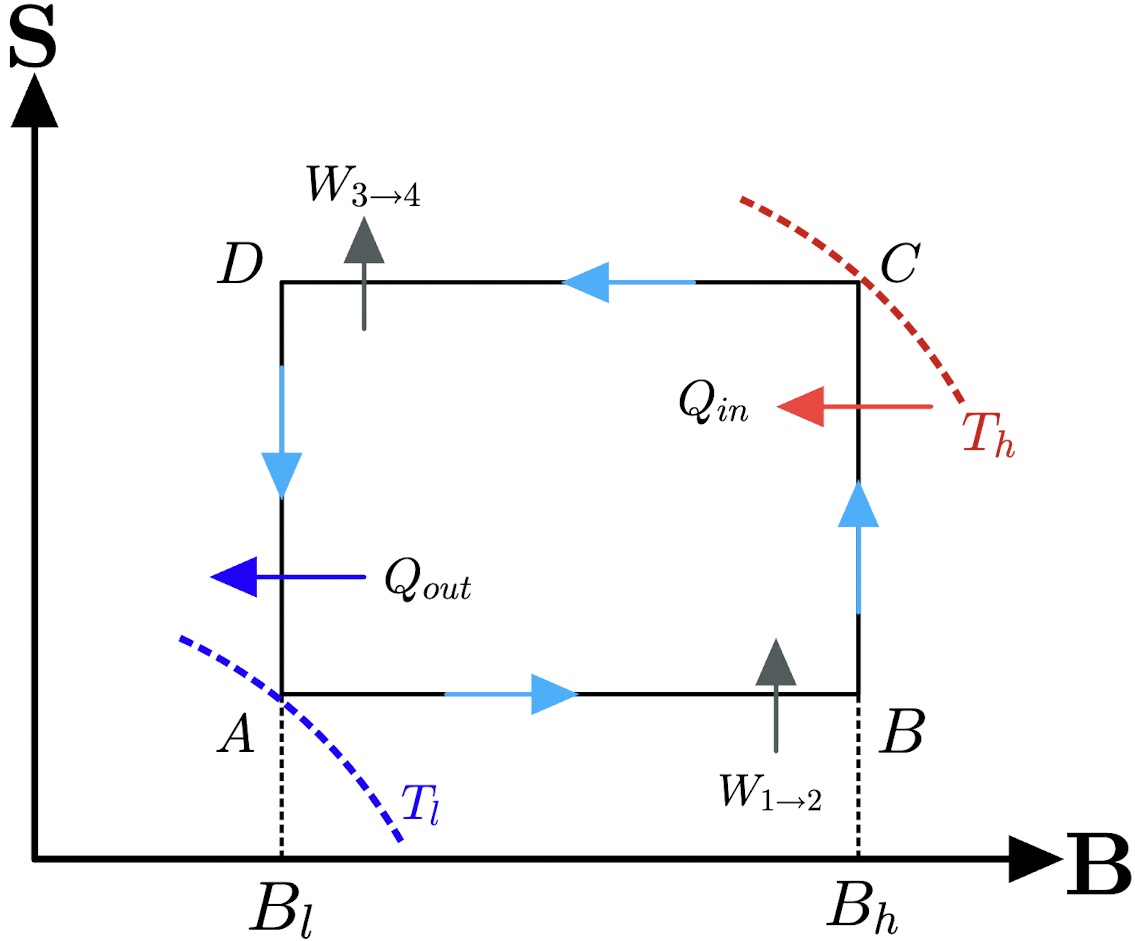}
\vspace{-4mm}
\caption{ \small Diagram of the Otto cycle on the plane of entropy and external magnetic field value along the z-axis. }
\label{ciclo_otto}
\end{figure}

Stage A-B:
In the cycle depicted in Fig. \ref{ciclo_otto}, the sequence \(\mathrm{A} \rightarrow \mathrm{B}\) starts with the system at an external magnetic field of \(B_l\), and thermal equilibrium at temperature \(T_l\). Subsequently, the magnetic field shifts from \(B_l\) to \(B_h\), elevating the internal temperature to \(T_B\) along an isoentropic path, thereby performing work.
\eq{
S(T_l,B_l)= S(T_B,B_h) &\implies &\nonumber \\ & W_{1 \rightarrow 2} &= U(T_B, B_h)- U(T_l,B_l).
}

Stage B-C:
The system is put in contact with a hot reservoir, increasing its internal temperature through heat absorption:
\eq{
Q_{in} = U(T_h,B_h)- U(T_B,B_h).
}

Stage C-D: The system is disconnected from the hot reservoir and modifies the magnetic field from \(B_h\) to \(B_l\) and traverses an additional isoentropic path, thus executing further work.
\eq{
S(T_D,B_l)= S(T_h,B_h) &\implies &\nonumber \\  & W_{3 \rightarrow 4} &= U(T_D, B_l)- U(T_h,B_h).
}

Stage D-A: Finally, the system is put in contact with a cold reservoir at temperature \(T_l\), expelling heat during the transition:
\eq{
Q_{out}=U(T_l,B_l)- U(T_D,B_l).
}
  
 The power output is quantified as the work conducted over each cycle iteration:
 \eq{
W = Q_{in} + Q_{out} .
 }
 Subsequently, the efficiency is defined as the proportion of total work relative to \(Q_{in}\):
 \eq{
 \eta = \frac{Q_{in}+Q_{out}}{Q_{in}}=1- \abs{\frac{Q_{out}}{Q_{in}}.} \label{eta_Otto_Eq}
 }
The resulting signs of work and {heat} in the different processes define the magnetic machine behavior. Although our main focus is on machines operating as engines, they could operate as refrigerators, heaters, and accelerators. The corresponding signs of heat and work in the different regimes are indicated in Table \ref{operation_modes}.
\begin{table}[ht]
    \centering
    \caption{Signs of heat and total work classifying different operational regimes.}
    \label{operation_modes}
    \begin{tabular}{|M{4cm}|M{1cm}|M{1cm}|M{1cm}|}
        \hline
         \textbf{Operational Regimes} & $W$ & $Q_{\text{in}}$ & $Q_{\text{out}}$ \\
        \hline
        Engine & $>0$ & $>0$ & $<0$ \\
        \hline
        Refrigerator & $<0$ & $<0$ & $>0$ \\
        \hline
        Heater & $<0$ & $<0$ & $<0$ \\
        \hline
        Accelerator & $<0$ & $>0$ & $<0$ \\
        \hline
    \end{tabular}
\end{table}

\section{Results and discussion}

In what follows we present results for systems with antiferromagnetic coupling, $J=1$, and different topologies $\alpha$.

\subsection{Stirling Engine }
{In this subsection, we will examine the impact of anisotropy on the efficiency of a Stirling cycle for temperatures below $T_c = 0.4$ in natural units of the model. The primary reason for this focus is that the cycle does not function as a thermal engine for $T>T_c$. Further discussion on this topic is provided below in this section.}

\begin{figure*} 
\begin{minipage}[b]{.45\textwidth}
\centering
\includegraphics[width=1\textwidth]{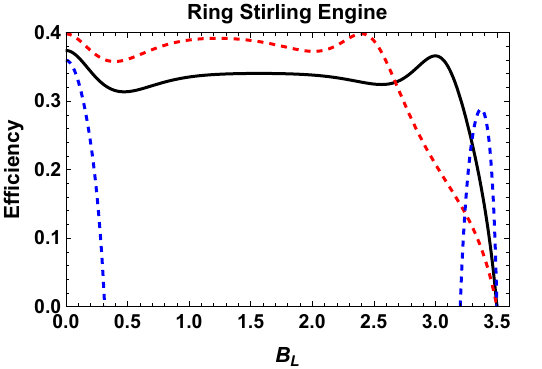}
\vspace{-8mm}
\caption{ Efficiency variation vs $B_L$ in ring ($\alpha=1$) antiferromagnetic ($J=1$) system with \(B_H=3.5\) and different \(K\) values. Cycle temperatures are \(T_L=0.2\) and \(T_H=0.33\). \(K=0\) (solid black); \(K=0.44\) (dashed blue); \(K=-0.44\) (dashed red).  Carnot efficiency \(\eta =0.4\) is reached for $K<0$ at the QCPs. }
\label{EtaStirlingEngineTri}
\end{minipage}
\hfill
\begin{minipage}[b]{.45\textwidth}
\centering
\includegraphics[width=1\textwidth]{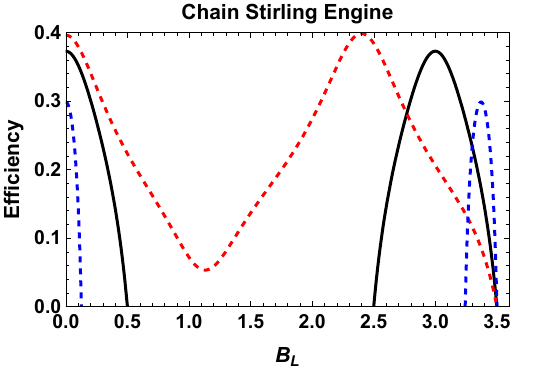}
\vspace{-8mm}
\caption{  Efficiency variation vs $B_L$ in chain ($\alpha=0$) antiferromagnetic ($J=1$) system with \(B_H=3.5\) and different \(K\) values.  Cycle temperatures are \(T_L=0.2\) and \(T_H=0.33\): \(K=0\) (solid black); \(K=0.44\) (dashed blue); \(K=-0.44\) (dashed red).  Carnot efficiency \(\eta =0.4\) is reached at QCPs for $K<0$.  }
\label{EtaStirlingEngineLine}
\end{minipage}
\end{figure*}

Figures \ref{EtaStirlingEngineTri}-\ref{EtaStirlingEngineLine} show the efficiency as a function of \(B_L\) for different values of anisotropy \(K = -0.44\) (red dashed line), \(K = 0\) (black thick line), and \(K = 0.44\) (blue dashed line), with fixed values of \(T_L = 0.2\), \(T_H = 0.33\), \(J = 1\), and different topologies \(\alpha = 1\) (ring) and \(\alpha = 0\) (chain).
We observe that the maximum efficiency for each \(K\) is located precisely at the quantum critical points (\(B_{\text{crit}} = -K + \sqrt{9J^2 + 12JK + K^2}\)). An important behavior is seen in both red dashed curves (\(K < 0\)). The ring (chain) system achieves Carnot efficiency at both critical points \(B = 0\) and \(B_{\text{crit}} \approx 2.48\). However, for positive \(K\), these values have lower efficiency and do not reach the Carnot values.  This may be related to the fact that for negative anisotropies, the energy levels are more closely spaced near the QCP, as shown in Fig.~\ref{PlotdeEnergias}(a)-(b), and more dispersed for positive anisotropies \cite{WOS:001130894000001}, as illustrated in Fig.~\ref{PlotdeEnergias}(c)-(d). 
This explains why the ring system exhibits overall higher efficiency than the chain system. Specifically, due to its degenerate nature, the triangular geometry has a higher number of states near the ground state between both QCPs at {all} temperatures. For systems with {negative} anisotropy, Carnot efficiency can be achieved at lower temperatures. 
In cases with no anisotropy, the efficiency is lower than negative anisotropy but higher overall than positive ones. 
\begin{figure}[t]
\centering
\includegraphics[width=.45\textwidth]{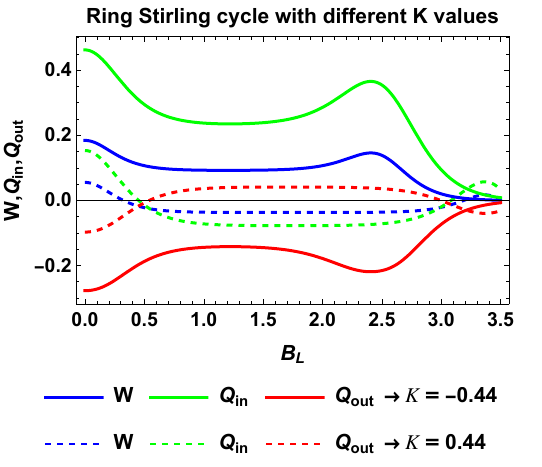}
\vspace{-4mm}
\caption{  {Work and heats for Stirling engine vs $B_L$ for antiferromagnetic ($J=1$) ring ($\alpha=1$) with different K values $K= -0.44$ (thick lines) and $K=0.44$ (dashed lines)  at fixed $T_L=0.2$, $T_H=0.33$, and $B_H=3.5$. The thick line in the cycle corresponds to the red dashed line in Fig.~\ref{EtaStirlingEngineTri}, while the dashed line cycle corresponds to the blue dashed line.  }}
\label{RingStirlingCycleK}
\end{figure}
  \begin{figure}[t] 
\centering
\includegraphics[width=.45\textwidth]{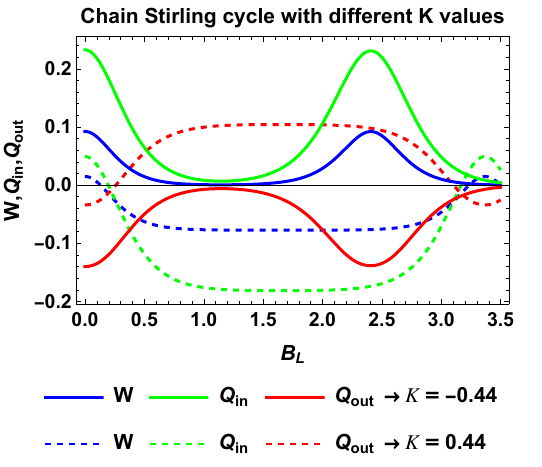}
\vspace{-4mm}
\caption{  { Work and heats for Stirling engine vs $B_L$ for antiferromagnetic ($J=1$) ring ($\alpha=1$) with different K values $K= -0.44$ (thick lines) and $K=0.44$ (dashed lines)  at fixed $T_L=0.2$, $T_H=0.33$, and $B_H=3.5$. 
The thick line cycle corresponds to the red dashed line in Fig.~\ref{EtaStirlingEngineLine}, while the dashed line cycle corresponds to the blue dashed line.}}
\label{ChainStirlingCycleK}
\end{figure}
\begin{figure*}[t]
\begin{minipage}[t]{.45\textwidth}
\centering
\includegraphics[width=.9\textwidth]{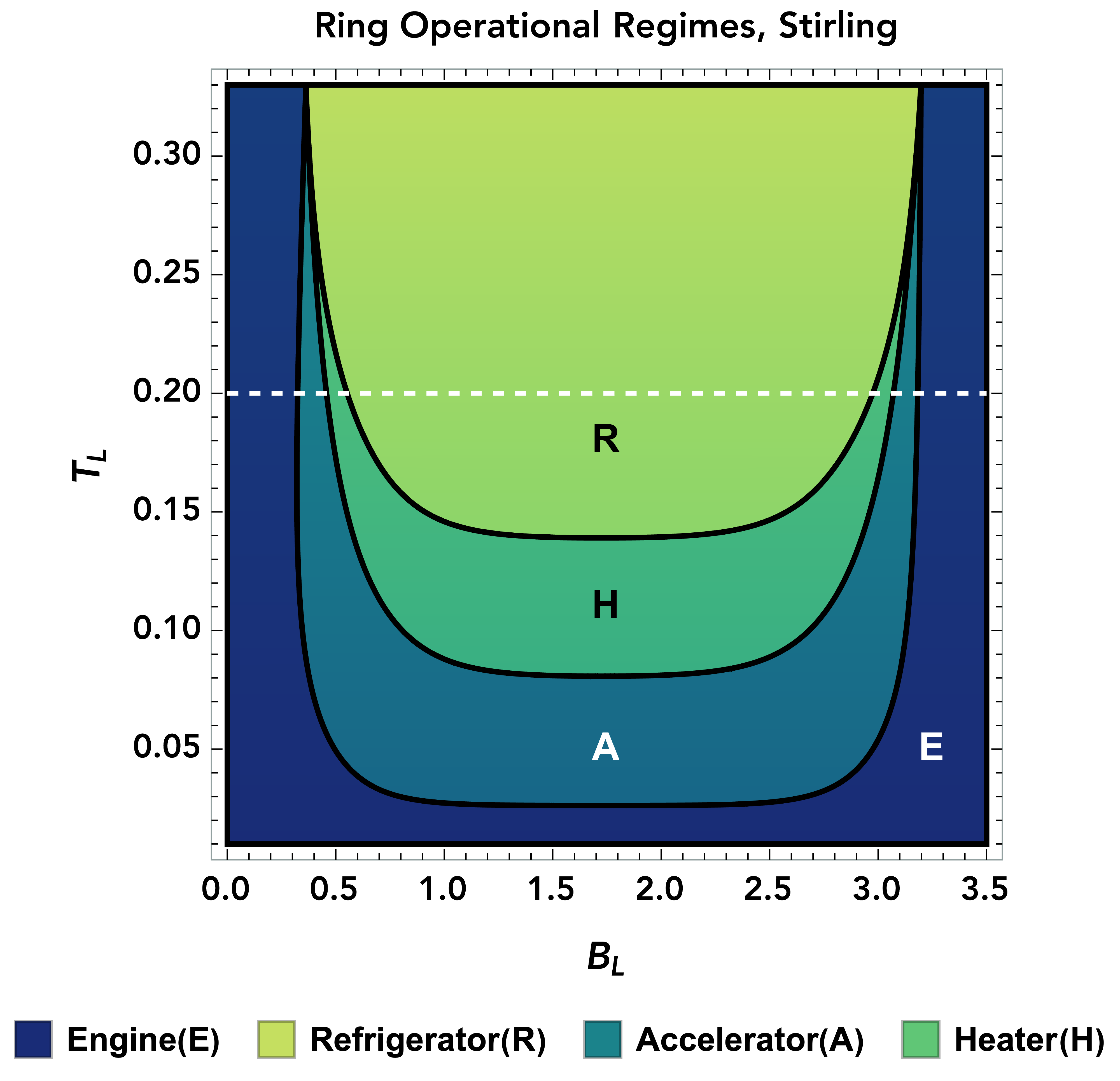}\vspace{-4mm}
\caption{ Stirling Engine operational regimes of the antiferromagnetic ($J=1$) ring ($\alpha=1$) with $K=0.44$ for different values of lower magnetic field $B_L$ and low temperature $T_L$ at fixed $B_H=3.5$ and $T_H=0.33$. The horizontal white dashed line represents the different operational regimes  { described by the dashed lines in Fig. \ref{RingStirlingCycleK}.}}
\label{operation_regimes_stirling}
\end{minipage}
\hfill
\begin{minipage}[t]{.45\textwidth}
\centering
\includegraphics[width=.9\textwidth]{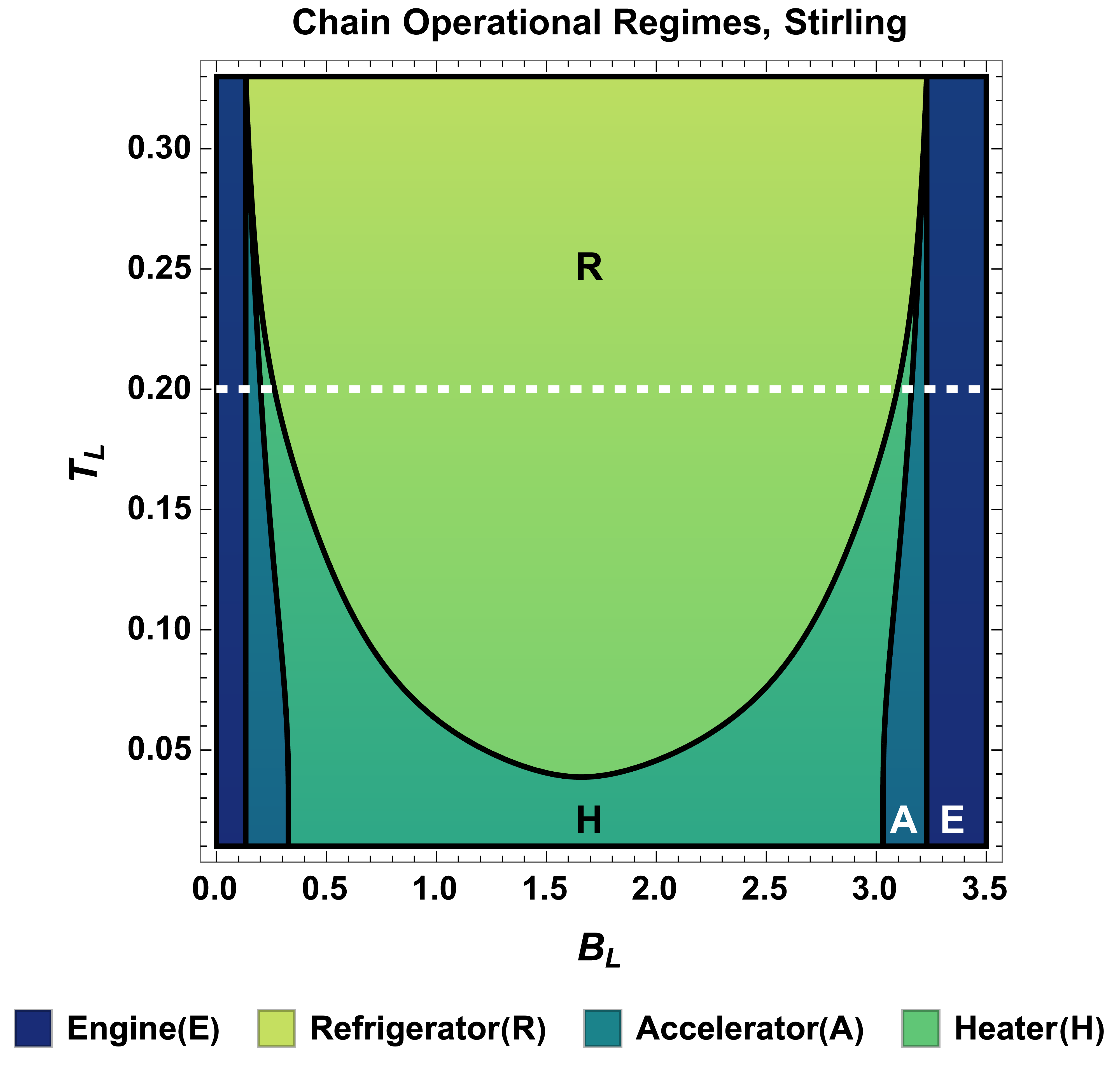}
\vspace{-4mm}
\caption{ Stirling Engine operational regimes of antiferromagnetic ($J=1$) chain ($\alpha=0$) with $K=0.44$ for different values of lower magnetic field $B_L$ and low temperature $T_L$ at fixed $B_H=3.5$ and $T_H=0.33$.  {The horizontal white dashed line represents the different operational regimes  described by the dashed lines in Fig. \ref{ChainStirlingCycleK}.}}
\label{operation_regimes_stirling_chain}
\end{minipage}
\end{figure*}
 \begin{figure} 
\centering
\includegraphics[width=.43\textwidth]{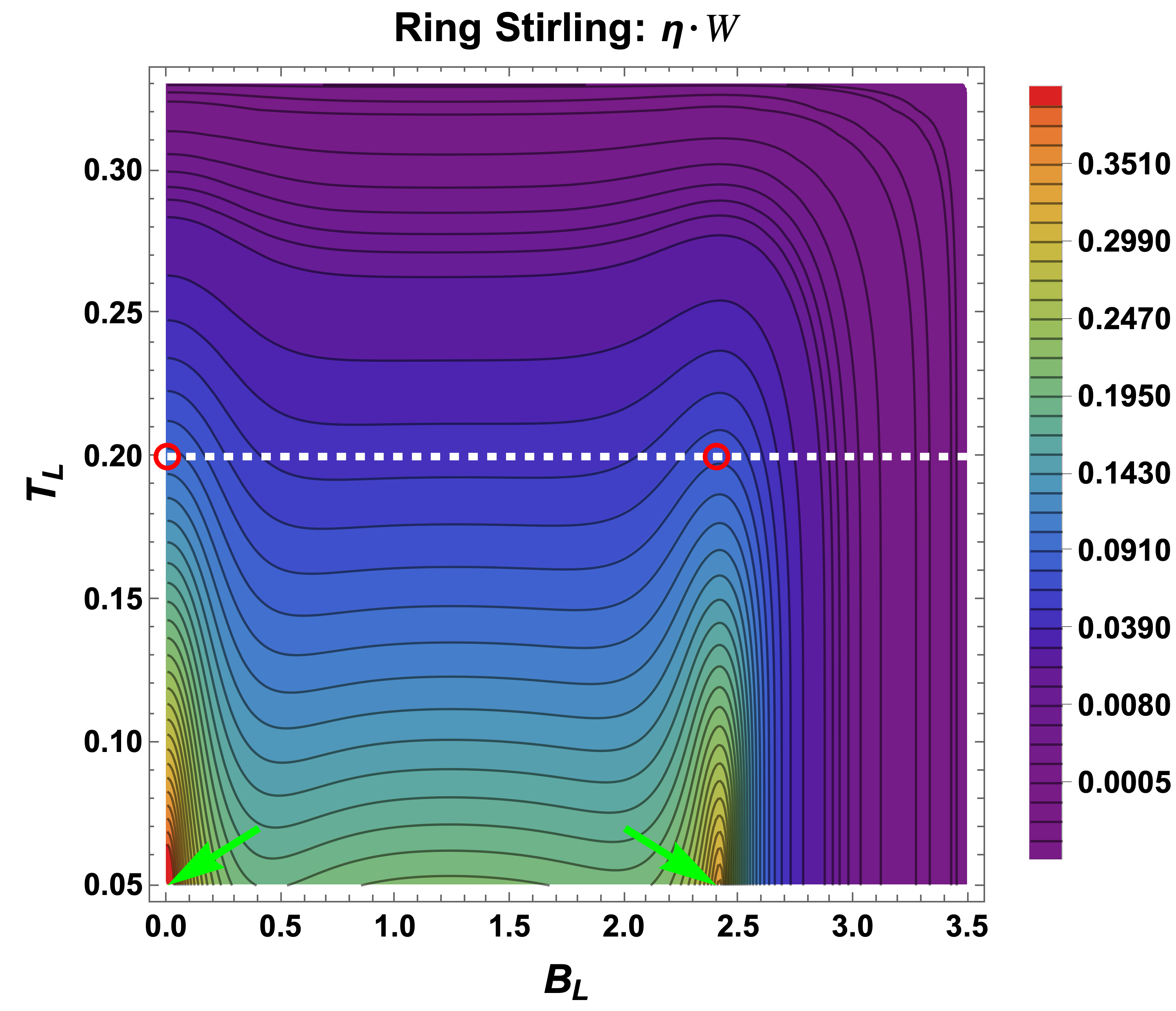}
\vspace{-4mm}
\caption{  Color map shows variation of the maximum gain, product of efficiency and work \((\eta \cdot W)\), for a ring (\(\alpha=1\)) antiferromagnetic system (\(J=1\)) as function of magnetic field \(B_L\) and temperature \(T_L\), with fixed \(T_H = 0.33\), $B_H=3.5$, and anisotropy \(K = -0.44\). This system operates as an engine across the full range of \(T_L\) and \(B_L\) shown.{White dashed line $(T_L=0.2) $ corresponds to the cycle depicted by the red dashed line in Fig.~\ref{EtaStirlingEngineTri}; red circles mark the points where this cycle achieves Carnot efficiency. Green arrows highlight the quantum critical point fields at $B_L=0$ and $B_L=2.4$. The bast purple regions on the color map signify areas where the scale is extremelly small but greater than zero. (not an absence of values.)}}
\label{etaWorkStirling}
\end{figure}
It is important to note that across all topologies, Carnot efficiency is never achieved with high-temperature reservoirs over than $T_c >0.4$. 
Another way to understand what happens with the system efficiency in a ring topology is to plot the behavior of the quantities $W$, $Q_{in}$ and $Q_{out}$ for the same parameter values as in  {Fig.~\ref{EtaStirlingEngineTri}. This is illustrated in Fig.~\ref{RingStirlingCycleK}, where the thick lines indicate a fixed anisotropy with $K = -0.44$. The maximum work output and heat input, along with the minimum heat output required for the system to function as an engine, are all located precisely at the Quantum Critical Point (QCP). At this point, the efficiency reaches its maximum, achieving the Carnot value on the ring topology. Conversely, the dashed lines in the same figure represent the cycle with a positive anisotropy value of $K = 0.44$. In this case, we also observe maximum and minimum values at the QCPs (shifting from $B=0$ to the right-shifted $B_{\rm crit}$) for the quantities $W$, $Q_{in}$, and $Q_{out}$. It is important to note, however the region where the system operates as an engine is significantly smaller and indicated by the blue dashed line in Fig.~\ref{EtaStirlingEngineTri}, the system does not achieve Carnot efficiency. 

A more pronounced behavior is observed in the chain topology for the quantities $W$, $Q_{in}$, and $Q_{out}$, with the same parameter values as shown in Fig.~\ref{EtaStirlingEngineLine}. For the same anisotropy values as before, the maximum work output and heat input, as well as the minimum heat output, are located at the QCPs. In this case, the system with negative anisotropy reaches Carnot efficiency, while the system with positive anisotropy does not. The intermediate values for the negative anisotropy cycle exhibit a notable minimum between the two QCPs than the ring topology, approaching zero but never quite reaching it. For positive anisotropy, the system behaves similarly to the ring topology, reaching nearly the same values at the second QCP and also does not achieve Carnot efficiency.. }

We emphasize that the Carnot efficiency for the cycle is attained through the utilization of a magnetic field, $B_L$, precisely at the quantum critical point (QCP). {This phenomenon arises due to the isomagnetic strokes undergoing simultaneous transformation into adiabatic strokes under the specified external parameters, which include anisotropy value, high and low temperatures, along with the high magnetic field strength, $B_H$.  {(See Appendix II. \ref{limiteffciency})}}
 We also note from Fig.~\ref{EtaStirlingEngineTri} and \ref{EtaStirlingEngineLine} that for $K>0$ both ring and chain topologies show that the system operates as engine over small and similar field ranges (blue dotted lines in both figures).  One can then conclude that both topologies with positive anisotropy will show similar work and efficiency. 
 
It is important to emphasize the signs of work and heat flows for negative anisotropy with the operation regimes in Table \ref{operation_modes}. As  {Figs.~\ref{RingStirlingCycleK}-\ref{ChainStirlingCycleK} for both topologies} demonstrates, throughout the entire range of \(B_L\), the  {systems} consistently operates as a heat engine, with positive \(W\) and \(Q_{in}\), and negative \(Q_{out}\). 
{However, the impact of topology is significant in the  {ring and chain systems} for {negative} and zero anisotropy cases. }

 \begin{figure}[h!]
\centering
\includegraphics[width=.45\textwidth]{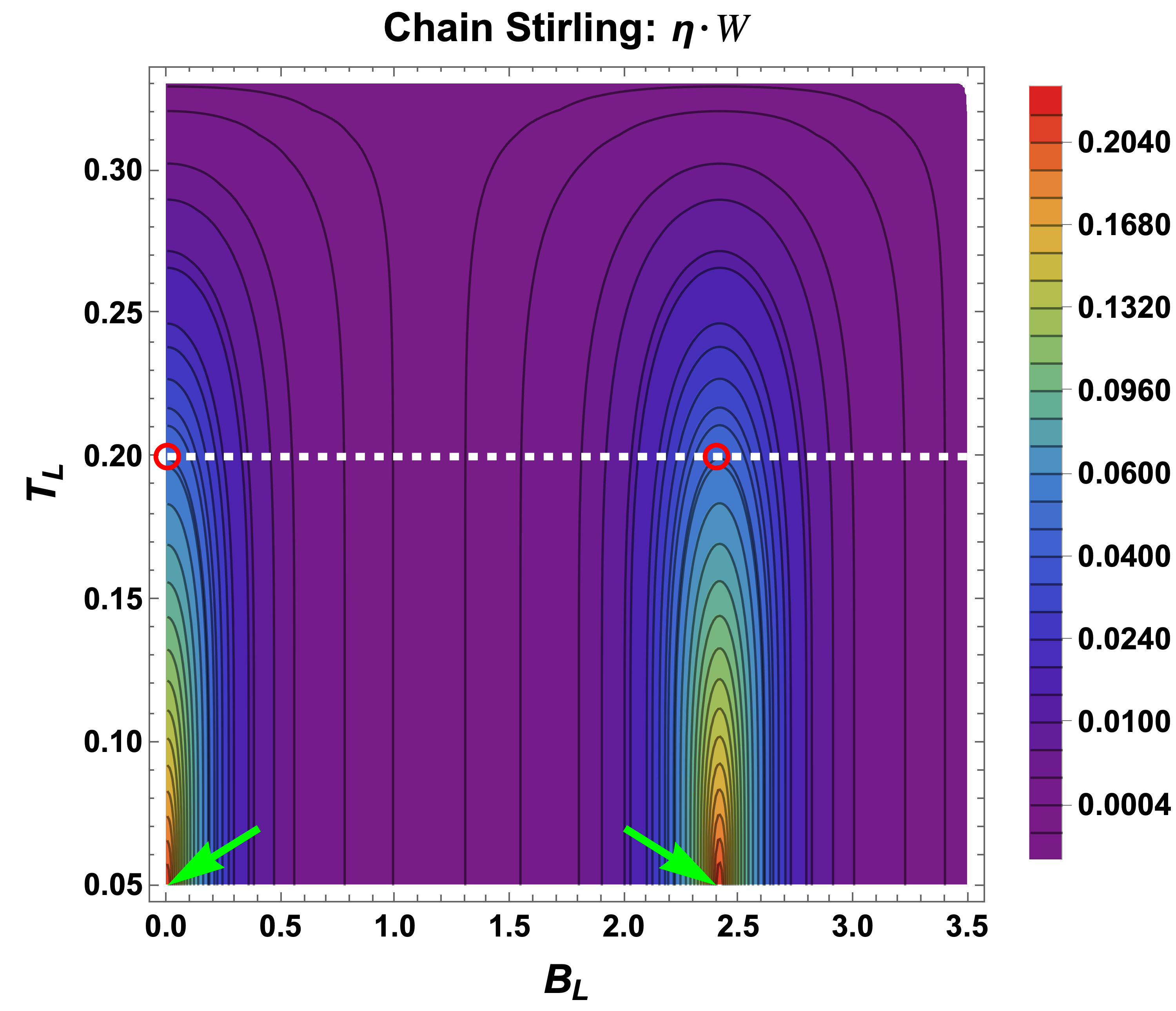}
\vspace{-4mm}
\caption{ Color map of maximum gain, efficiency times work \((\eta \cdot W)\) for an chain ($\alpha=0$) antiferromagnetic ($J=1$) system vs magnetic field \(B_L\) and temperature \(T_L\), with fixed $T_H=0.33$, $B_H=3.5$, and \(K = -0.44\). The system behaves as an engine across full range of \(T_L\) and \(B_L\). { White dashed line $(T_L=0.2) $ corresponds to the cycle depicted by the red dashed line in Fig.~\ref{EtaStirlingEngineLine}; red circles mark points where this cycle achieves Carnot efficiency. Green arrows highlight the locations of the quantum critical fields. The bast purple regions on the color map signify areas where the scale is extremelly small but greater than zero. (not an absence of values.)} }
\label{etaWorkStirlingChain}
\end{figure}
 The  {systems} with negative anisotropy behaves as an engine across all magnetic field values, whereas the same system with positive anisotropy does not. it only exhibits engine properties at low and high \(B_L\) values, with \(W, Q_{in} > 0\) and \(Q_{out} < 0\) as mentioned previously. At intermediate values  {for the positive anisotropy, the systems transitioned} to a refrigerator. More notably, the transition between an engine and a refrigerator is marked by brief intervals through the heater and accelerator regimes. It begins with an engine transitioning to an accelerator, and then for a brief interval, it becomes a heater as \(Q_{out}\) shifts from negative to positive and $Q_{in}$ from positive to negative, subsequently transforming into a refrigerator for almost the entire range of the magnetic field. Subsequently, it cycles through a heater regime again before briefly returning to an accelerator. Ultimately, it transitions back to an engine. 

To characterize the different Stirling machine behaviors exhibited when $K > 0$, we study systems with fixed parameters $B_{H}=3.5$, $T_{H}=0.33$ and $K=0.44$ for a ring topology ($\alpha=1$) in Fig.~\ref{operation_regimes_stirling} and for an chain topology ($\alpha=0$) in Fig.~\ref{operation_regimes_stirling_chain}. In both graphics we sweep $B_{L}$ from 0 to $B_{H}$ and $T_{L}$ from 0.01 to $T_{H}$ and classify the different operational regimes.
In the case of the ring topology, we can observe from Fig.~\ref{operation_regimes_stirling} that at very low temperatures $(T_{L} < 0.03)$, there is always the possibility of finding an operational engine behavior for the entire range of $B_{L}$ studied.  The same behavior occurs for higher $T_{L}$ if we confine operation at  low $(B_{L} < 0.5)$ or high $(B_{L} > 3.1)$ fields. Notice that if we set $T_{L} = 0.20$, indicated by a horizontal white dashed line, the system goes through four regimes, dominated by the refrigerator behavior at intermediate fields, $0.5 \lesssim B_L \lesssim 2.9$. For the case of the chain topology presented in Fig.~\ref{operation_regimes_stirling_chain}, the regimes differ mainly in that there is no operational engine behavior for any $B_{L}$ in this region. The system presents sharp transitions to narrow regions where the accelerator regime is seen. The regime that increases to compensate and dominates this range corresponds to the refrigerator operation.

\begin{figure}
\centering
\begin{minipage}[b]{.44\textwidth}
\centering
\includegraphics[width=1\textwidth]{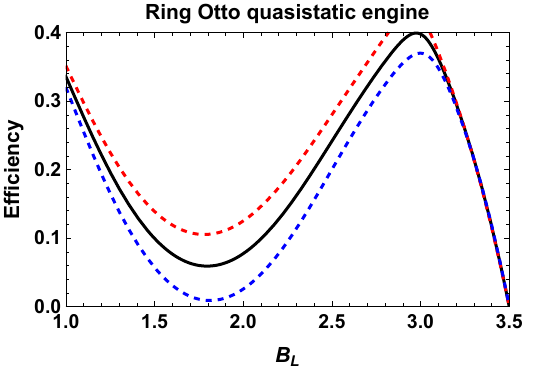}
\caption{ Efficiency variation with $B_L$ in ring ($\alpha=1$) antiferromagnetic $(J=1)$ system at \(B_H=3.5\), \(T_L=0.94\), and \(T_H=1.56\) for different \(K\) values: \(K=0\) (solid black); \( K=0.033\) (dashed blue); \( K=-0.033\) (dashed red); Carnot efficiency is \(\eta =0.4\). }
\label{EtaOttoQuasistaticTri}
\end{minipage}
\end{figure}
\begin{figure} 
\begin{minipage}[b]{.44\textwidth}
\centering
\includegraphics[width=1\textwidth]{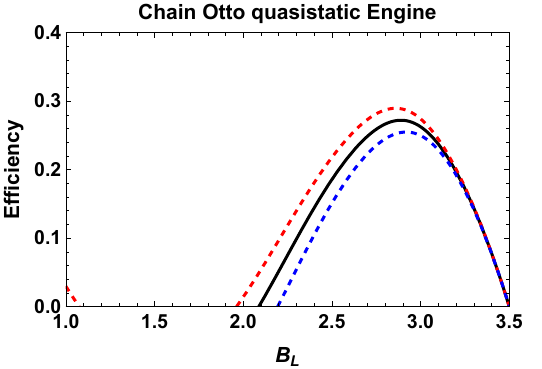}
\caption{  Efficiency variation with $B_L$ in chain ($\alpha=0$) antiferromagnetic $(J=1)$ system at \(B_H=3.5\), \(T_L=0.94\), and \(T_H=1.56\) for different \(K\) values: \(K=0\) (solid black); \(K=0.033\) (dashed blue); \(K=-0.033\) (dashed red); \ Carnot efficiency is \(\eta =0.4\). }
\label{EtaOttoQuasistaticLinea}
\end{minipage}
\end{figure}

  An interesting aspect to consider is whether maximum efficiency corresponds to maximum output. Fig.~\ref{etaWorkStirling} shows the maximum gain, given by the product of the work output and the efficiency, $\eta \cdot W$, for a ring system with negative anisotropy.  The Carnot efficiency, red dashed line in Fig.~\ref{EtaStirlingEngineTri}, occurs at the local maximum gain \(\eta \cdot W\). It is also noteworthy that the optimal operation conditions for maximum gains occur near the QCPs at low temperatures.
 We note also that after the second QCP of the model the maximum gain $\eta \cdot W$ is essentially constant independent of $B_L$, as shown by the horizontal contour lines. 
 \begin{figure}[h]
    \begin{minipage}[b]{.45\textwidth}
\centering
\includegraphics[width=1\textwidth]{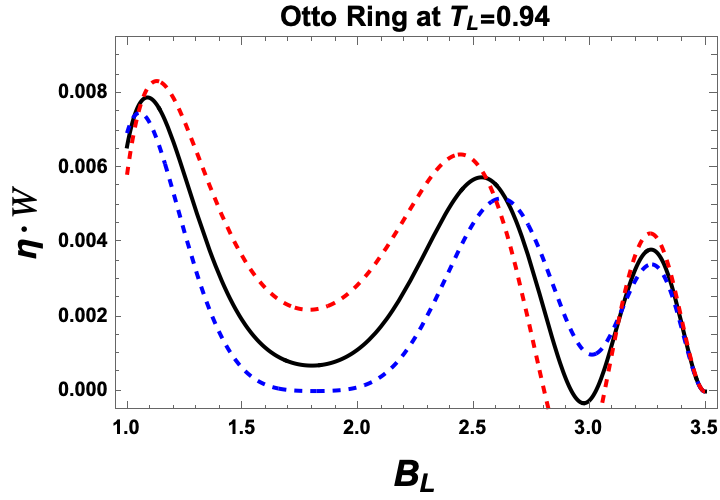}
\caption{ Otto quasistatic maximum gain $\eta \cdot W $ as function of the lower magnetic field $B_L$ for an antiferromagnetic $(J=1)$ ring ($\alpha=1$) at different values of anisotropy: $K=0$ (black solid), $K=0.033$ (blue dashed) and $K=-0.033$ (red dashed) at $T_L = 0.94$ and $T_H=1.56$. }
\label{GainsOttoQuasis1}
\end{minipage}
\end{figure}
\begin{figure}  
\begin{minipage}{.45\textwidth}
\centering
\includegraphics[width=.95\textwidth]{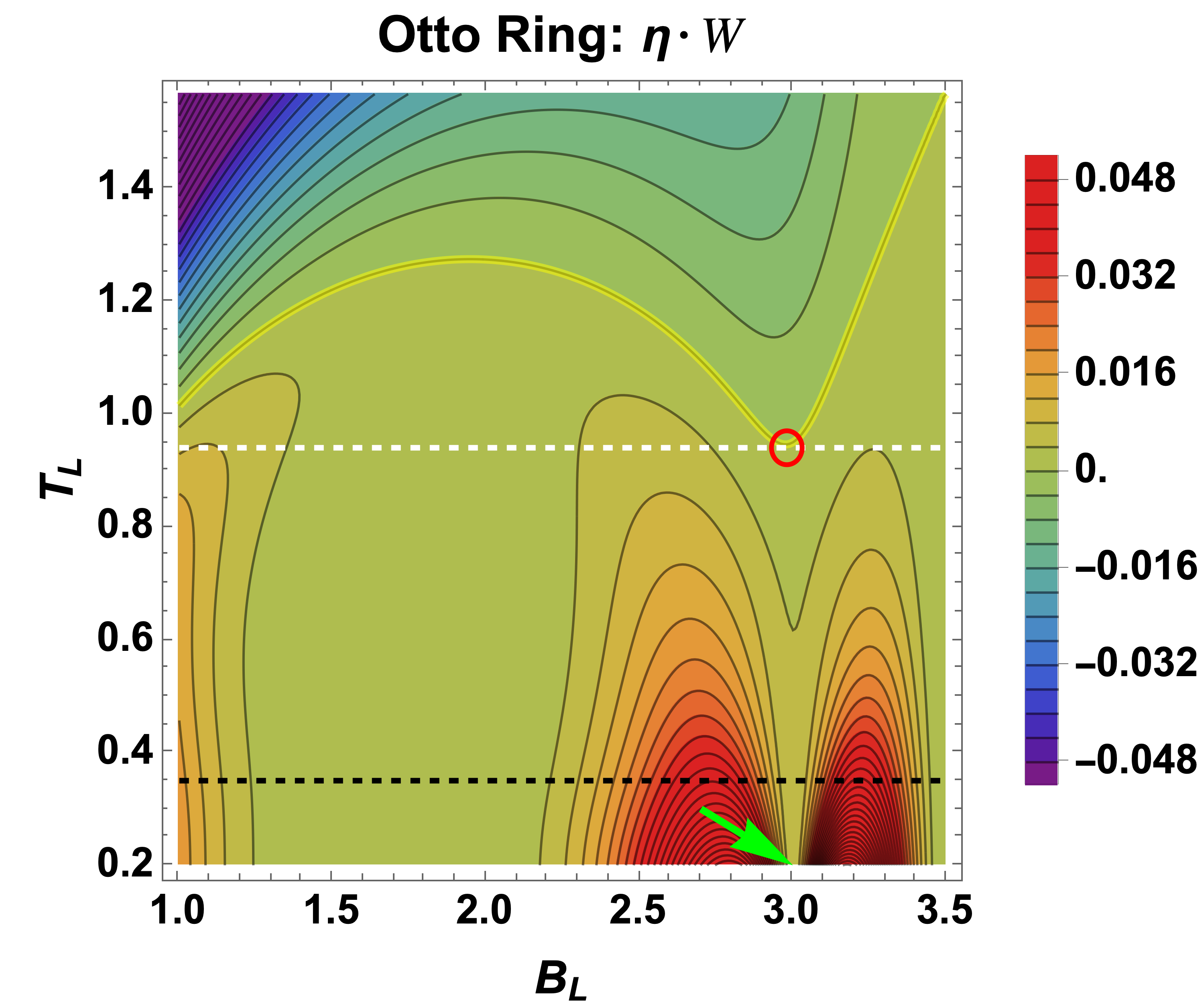}
\vspace{-4mm}
\caption{  Color map of maximum gain, efficiency times work \((\eta \cdot W)\) in an antiferromagnetic ring \((\alpha=1, J=1)\) vs \(B_L\) and \(T_L\) with fixed  {$T_H=1.56$}, $B_H=3.5$, and \(K = 0\). The yellow line indicates where the function equals zero; the system operates as a refrigerator above this line and as an engine below it. {A dashed white line highlights the cycle shown by the black line in Fig.~\ref{EtaOttoQuasistaticTri}. The horizontal dashed black line represents the zero anisotropy cycle in Fig.~\ref{GainsOttoQuasisLowT}. At high temperatures, the zones become extremely negative, whereas near the QCP, they exhibit exceptionally high positive values. A green arrow points to the QCP field.} }
\label{GainsOtto}
\end{minipage}
\end{figure}

  \begin{figure}
    \begin{minipage}{.45\textwidth}
\centering
\includegraphics[width=1\textwidth]{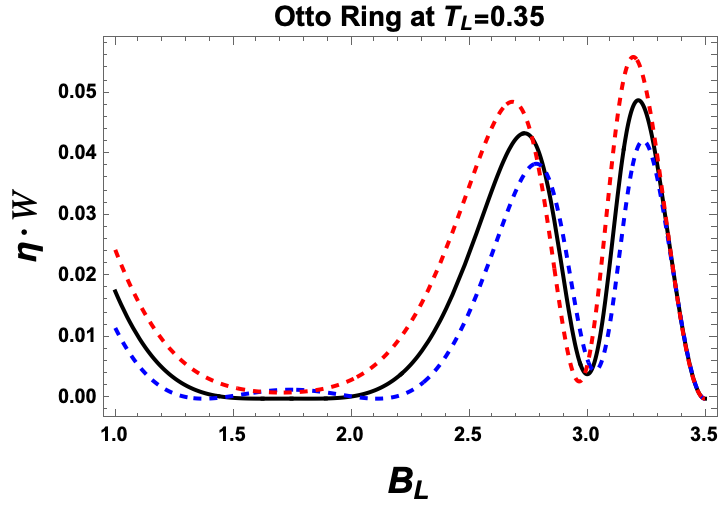}
\caption{ Otto quasistatic maximum gain $\eta \cdot  W $ as function of the lower magnetic field $B_L$ for an antiferromagnetic $(J=1)$ ring ($\alpha=1$) at different values of anisotropy: $K=0$ (black), $K=0.033$ (blue) and $K=-0.033$ (red) at $T_L = 0.35$ and $T_H=1.56$. }
\label{GainsOttoQuasisLowT}
\end{minipage}
\end{figure} 
It is evident that topology influences performance. The ring system achieves better results than the chain system at both low and medium temperatures. As the temperature increases, leading to a greater population of higher states, this tends to affect efficiency negatively, preventing the attainment of Carnot efficiency. Notably, the systems that reach Carnot efficiency typically populate only two levels: the ground state and the first excited state. Fig.~\ref{etaWorkStirlingChain} shows maximum gain for an chain structure with negative anisotropy.  Notice the overall scale of the map is lower than in the ring system, although ridges of larger gain also appear near the QCPs.  In contrast to the ring topology, for $B_L$ beyond the QCP, vertical lines in the contour plot indicate $T_L$-independent gain values.
  

\subsection{Quasistatic Otto Engine}
In this subsection, we delve into the behavior of the quasistatic Otto cycle. While there are some similarities to the Stirling engine, the results from the Otto cycle present notable differences. Unlike the Stirling cycle, {the Otto cycle operates as an engine only at high temperatures and magnetic fields, specifically above $T\sim  0.9$ and $B_{L} > 1$.}
Fig.~\ref{EtaOttoQuasistaticTri} illustrates that the efficiency of the quasistatic Otto cycle in the ring system achieves Carnot efficiency at zero anisotropy (depicted by the black line) and operates as an engine across the entire magnetic field range. For positive anisotropies, efficiency increases, while for negative anisotropies, it decreases. When the efficiency exceeds Carnot's, the system functions as a refrigerator. Conversely, if the efficiency drops below zero, it also transitions to a refrigeration mode.
In Fig.~\ref{EtaOttoQuasistaticLinea}, the chain system's behavior is distinct. It predominantly operates as a refrigerator at low values of the magnetic field \(B_L\) and switches to an engine mode at higher  field. The efficiencies of the chain system consistently lag behind those of the ring system and never achieve Carnot efficiency. 
This contrasts with the Stirling engine's capability to transition through all thermal operational regimes. In comparison, the quasistatic Otto cycle is constrained to operate as either an engine or a refrigerator.
This distinction between the two thermodynamic cycles underscores a notable difference in their operational versatility.

In the context of the thermodynamic cycle under investigation, a salient question emerges: does anisotropy increase or decrease the maximum possible work output? In Fig.~\ref{GainsOttoQuasis1}, we observe that across nearly the entire magnetic field range, negative anisotropy enhances the \(\eta \cdot W\) product except near \(B_L = 3.0\), where the system transitions to a refrigerator. Conversely, positive anisotropy reduces the \(\eta \cdot W\) product for \(B_L < 2.7\); near this point, the system outperforms the zero anisotropy configuration. In Fig.~\ref{GainsOtto}, we present a contour map of \(\eta \cdot W\) for the quasistatic Otto cycle vs \(T_L\) and \(B_L\) at zero anisotropy, $K=0$. The yellow curve represents Carnot efficiency but \(\eta \cdot W = 0\). This represents a significant limit of a quasi-static thermodynamic cycle of the Otto type \cite{pena1}, where the work approaches zero as both \( Q_H \) and \( Q_L \) approach zero. As indicated by Eq~\eqref{eta_Otto_Eq}, the limit ratio takes an indeterminate form, but both heats approach the limit \( T_L / T_H \).
{One can achieve Carnot efficiency at the local minimum of this line, as indicated by the red circle marking the field value at the QCP. The system operates entirely as an engine across the entire magnetic field range, as indicated by the green arrows.}

However, this is not the most advantageous location for maximum gains ($\eta \cdot W $), as these occur near the quantum critical points at low temperatures, as indicated by the horizontal black line in Fig.~\ref{GainsOtto}. It is useful to plot the results for a cycle in this region. Fig.~\ref{GainsOttoQuasisLowT} does this for different anisotropy values. The results are indeed significantly more favorable to thoe in Fig.~\ref{GainsOttoQuasis1}, with this cycle having
a nearly sixfold enhancement in performance near the QCP at \(B_L = 3.0\). It should be noticed that unlike the Stirling cycle, the quasistatic Otto cycle performs minimally at the QCP.

\section{Conclusions}
{We have studied a three-qubit system with antiferromagnetic exchange interactions between nearest neighbors, subjected to a homogeneous magnetic field along the z-direction and anisotropy along the y-direction. We observed significant improvements in engine efficiency with negative (easy-axis) anisotropy across all thermodynamic cycles, regardless of the ring or chain topology.}

We find that the choice of topology markedly influences the behavior of the Stirling cycle. Specifically, at lower temperatures, work output and efficiency are enhanced in the ring topology compared to the open chain. These trends  extend to the quasistatic Otto cycle, albeit with lower maximum gains $\eta \cdot W$ at elevated temperatures than in the Stirling cycle.

In the Stirling cycle and for negative anisotropy, the system reaches Carnot efficiency in both topologies while producing finite work.  
{The Carnot efficiency for the cycle is attained through the utilization of the low magnetic field, precisely at the Quantum Critical Point (QCP). This phenomenon arises due to the isomagnetic strokes undergoing simultaneous transformation into adiabatic strokes under the specified external parameters, which include anisotropy value, high and low temperatures, and high magnetic field. Therefore transforming the Stirling cycle into a Carnot cycle.}

On the contrary, the quasistatic Otto engine reaches Carnot efficiency at the QCPs but fails to generate useful work. The latter is a sign of the robustness of thermodynamics as it is an expected limit when working with quasi-static Otto cycles independent of the nature of the working substance.

Notably, the Stirling cycle exhibits versatility across all thermal operational regimes: as an engine, refrigerator, heater, and accelerator. Leveraging this property, one could envision utilizing the Stirling cycle to cool a three-qubit configuration, effectively acting as a heater system.

 The average work performed by this 3 qubits system is of {$W=0.1J$ for the linear chain and 
to $W=0.2J$} for a triangle, both cases for a Stirling cycle. In the case of the Otto cycle, these values are $W=0.01J$ and $W=0.03J$, respectively.  Entropy values of about $S = 0$ to $S=k_B \log(3)$
are found in both topologies, {in line with one-level to three-level systems}, which are the levels that are thermally populated in both cycles.

\section{Acknowledgments}
B.C, F.J.P., C.A, S.E.U. and P.V. acknowledge financial support from ANID
Fondecyt grant no. 1240582. F.J.P. acknowledges “Millennium Nucleus
in NanoBioPhysics” project NNBP NCN2021 021. B.C acknowledge PUCV and ''Direcci\'on de Postgrado'' of UTFSM.

\section{Appendix I}
\subsection{Variable Definition}\label{variable_change}
The variable use on the article are, respectively
\eq{P_\pm =\sqrt{B^2 \pm 2 BK +4 K^2}  } 
\eq{R_{i}^\pm =\pm B+ K +(-1)^i P_\pm }
\eq{L_{i}^\pm = \dfrac{1}{\sqrt{3+ \qty(\frac{ R_{i}^\pm }{K})^2}}}

\eq{J_{\text{eff}}  = J (2+\alpha) + 5 K}
\subsection{Hamiltonian Matrix Representation}
{We use  {the standard} basis to write the matrix $\sigma_z$ representation using  { $\ket{0}=[1,0]^T$ and $\ket{1}=[0,1]^T$}. Therefore for three-qubit basis:  { $\ket{000},\ket{001},\ket{010},\ket{011},\ket{100},\ket{101},\ket{110},\ket{111} $}. The Hamiltonian \eqref{H} has the form: }
\begin{widetext}
\eq{
\mathcal{H} = \rescale[1]{\mqty(
3(K+B)+J(2+ \alpha)  & 0 & 0 &-2K &0 & -2K & -2K &0\\
0 & B- J \alpha+3K & 2(J +K) & 0 & 2(J \alpha+K) & 0 & 0 &-2K\\
0& 2(J+K)& B+ J(\alpha-2)+3K & 0 & 2(J+K) &0 & 0& -2K\\
-2K & 0 &0 & -B - J \alpha + 3K &0 &2(J+ K) & 2(J \alpha + K) &0 \\
0& 2 (J \alpha +K) & 2(J+K) & 0& B -J \alpha +3 K  & 0 & 0 & - 2K \\
-2K & 0 & 0 & 2(J+K) & 0 & -B + J(\alpha -2) +3 K & 2(J+K) & 0 \\
-2K & 0 & 0& 2(J \alpha+K) & 0 & 2(J+K) & -B - J\alpha +3K &0 \\
0 & - 2K & -2K &0 & -2K &0 &0 & 3(K- B) +J(\alpha +2) 
)}
}
\end{widetext} 
 \subsection{Experimental Viability}
 Experiments to determine the exchange coupling \(J\) values for interacting qubits have been reported in recent literature \cite{WOS:000501493200001, datos_2005}. Recent experiment \cite{datos_2021} on a linear chain of three qubits made of quantum dots in Silicon establish that the nearest-neighbor exchange constant is approximately \(J \approx h (40 \, \text{GHz}) \approx 160 \, \mu\text{eV}\),
 and the superexchange constant between the chain's extremes is \(J_{\text{exp}} \approx h (6.5 \, \text{MHz}) \sim 0.02 \, \mu\text{eV}\), which is effectively negligible in this context.
 Another recent experiment \cite{nature_ariel} done on two holes spin qubits in Silicon thin field effect transistor reports an exchange coupling of $h(90 \text{MHz})$ corresponding to a $J=0.4\, \mu eV$ value.
 Both experiments were done at temperatures of about $40 mK$.
 For the Quantum Dot data in our case, these results indicate an average work of { $W\sim 16\, \mu eV $ for the linear chain and to $W \sim 32\, \mu eV $ for a triangle, both cases for a Stirling cycle.} In the case of the Otto cycle, these values are $W \sim 1.6 \,\mu eV$ and $W \sim 4.8\, \mu eV$, respectively.  
For the quantum hole data, an average work of {$W \sim 0.04 \,\mu eV $ for the linear chain and $W \sim 0.08 \, \mu eV $ for a ring}, both cases for a Stirling cycle. In the case of the Otto cycle, these values are $W \sim 0.012\, \mu eV$ and $W\sim 0.004 \,\mu eV$, respectively.  

 {\section{Appendix II}
\subsection{Carnot into Stirling cycle}\label{limiteffciency}}

 {
{In general, the efficiency of the Stirling cycle of Fig.~\ref{ciclo_stirling} can be written in the form }

{
\eq{
\label{etageneral}
    \eta= & \rescale[.45]{ \dfrac{T_H \qty(S_B-S_A) + T_L(S_D-S_C) + (U_C-U_B)+ (U_D- U_A)}{T_H (S_B - S_A) + (U_A-U_D)} }
}}

{At low temperatures, we can think of analyzing this cycle as an effective system of two generic energy levels depending on the control parameter $B$ (in our case, the external magnetic field) $\varepsilon_{1}$ and $\varepsilon_{2}$, which have different degeneracy given by $g_{1}$ and $g_{2}$ (with $g_{2} \geq g_{1}$). When varying the external field a QCP is found at $B=B_{\text{crit}}$ as shown in Fig.~\ref{appendix_2_fig_1}, with a difference between energies at $B=B_{H}$ that satisfies $\varepsilon_{1}-\varepsilon_{2}=\Delta$.}

\begin{figure}
    \centering
    \includegraphics[width=.7\linewidth]{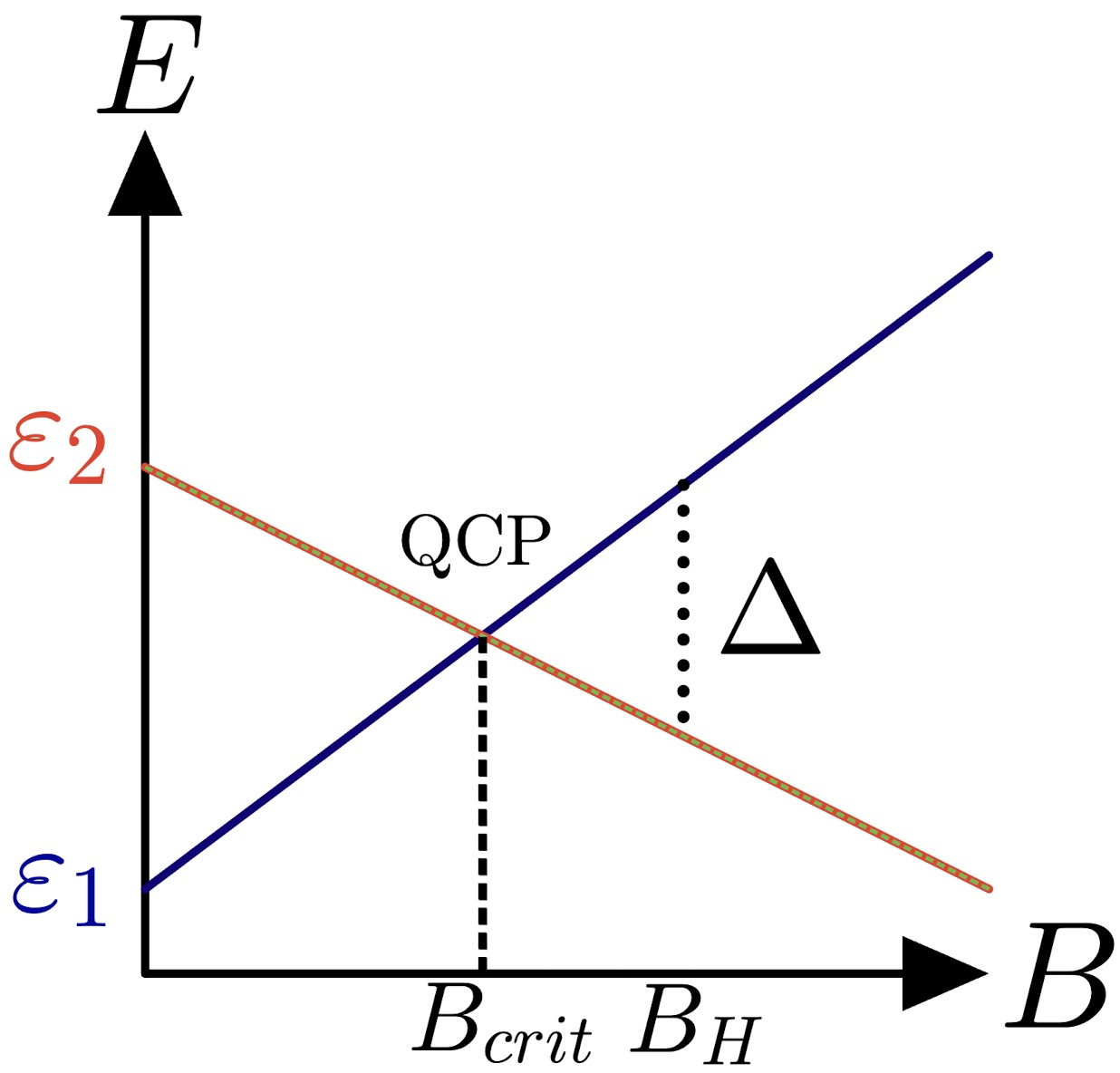}
    \caption{ { Ground state $\varepsilon_1$ and first excited state $\varepsilon_2$ plotted against the external magnetic field $B$. The cross point is the Quantum Critical Point at $B_{\text{crit}}$, and $\Delta$ represents the separation of energies.}}
    \label{appendix_2_fig_1}
\end{figure}

{The conditions that we will impose in executing the cycle are the following: First, at $B=B_{H}$ when $B_H > B_{crit}$, the system will be in the fundamental state with energy $\varepsilon_{2}$, and at $B=B_{L}=B_{crit}$, the system presents a QCP because the ground of the system has changed as we changed the external magnetic field in the execution of the cycle. Also, the following analysis assumes that the temperature of the reservoirs is close and low enough, so that $(T_{H}- T_L)/T_{H} \sim \delta \ll  1$, and only the ground state of the system is populated during the cycle.} 

{If we focus our study on the isomagnetic stages of the thermodynamic cycle of Fig.~\ref{ciclo_stirling}, for the process $D\rightarrow A$, and only the ground energy state $\varepsilon_{2}$ is populated due to the large energy gap $(\Delta)$ at low temperature, we will have that the internal energies of the system satisfy $U_{A}(T_{H},B_{H})\approx U_{D}(T_{L},B_{L})$. In the same way, the entropy at these points will satisfy $S_{A}(T_{H},B_{H})\approx S_{D}(T_{L},B_{L}) \sim\ln\left(g_{2}\right)$ due to the small difference between the reservoirs. In the second isomagnetic stage, $B \rightarrow C$, a process that takes place precisely where the system presents a QCP on $B_L = B_{\text{crit}}$, we have a change in the ground state of the system that has both energies at the same point $\varepsilon_{1}=\varepsilon_{2}$. Consequently, the internal energies in the stage are the same, which means $U_{C}(T_{L},B_{\text{crit}}) =  U_{B}(T_{H}, B_{\text{crit}})= {\rm const}$. If the difference between temperatures is low enough $\delta \ll 1$, it implies that the entropy in both points is the same, which means  $S_{C}(T_{L},B_{\text{crit}})\approx S_{B}(T_{H},B_{\text{crit}}) \sim\ln\left(g_{1}\right)+ \ln \qty(g_2) = \ln\qty(g_1 g_2)$. These approximations imply that the Stirling cycle, which originally has two isothermal and two isomagnetic paths, now consists of two isothermal and two essentially adiabatic paths, effectively transforming it into a Carnot cycle. This can be seen replacing these approximate relations into Eq.~(\ref{etageneral}), as we obtain }

\begin{align} 
  \begin{aligned}
        \eta &\approx 1 + \frac{T_L}{T_H}\cdot  \frac{S_D(T_L, B_H) - S_C(T_L, B_{\text{crit}})}{S_B(T_H, B_{\text{crit}}) - S_A(T_H, B_H)} \\
    &\approx 1 + \frac{T_L}{T_H}\cdot  \frac{\ln \qty(g_2) - \ln \qty(g_1 g_2)}{\ln \qty(g_1 g_2)- \ln \qty(g_2)} \\
    & \approx 1 - \frac{T_L}{T_H} \, .
  \end{aligned}\label{etageneral2}
\end{align}
 \begin{figure}[t]
    \centering
    \includegraphics[width=1\linewidth]{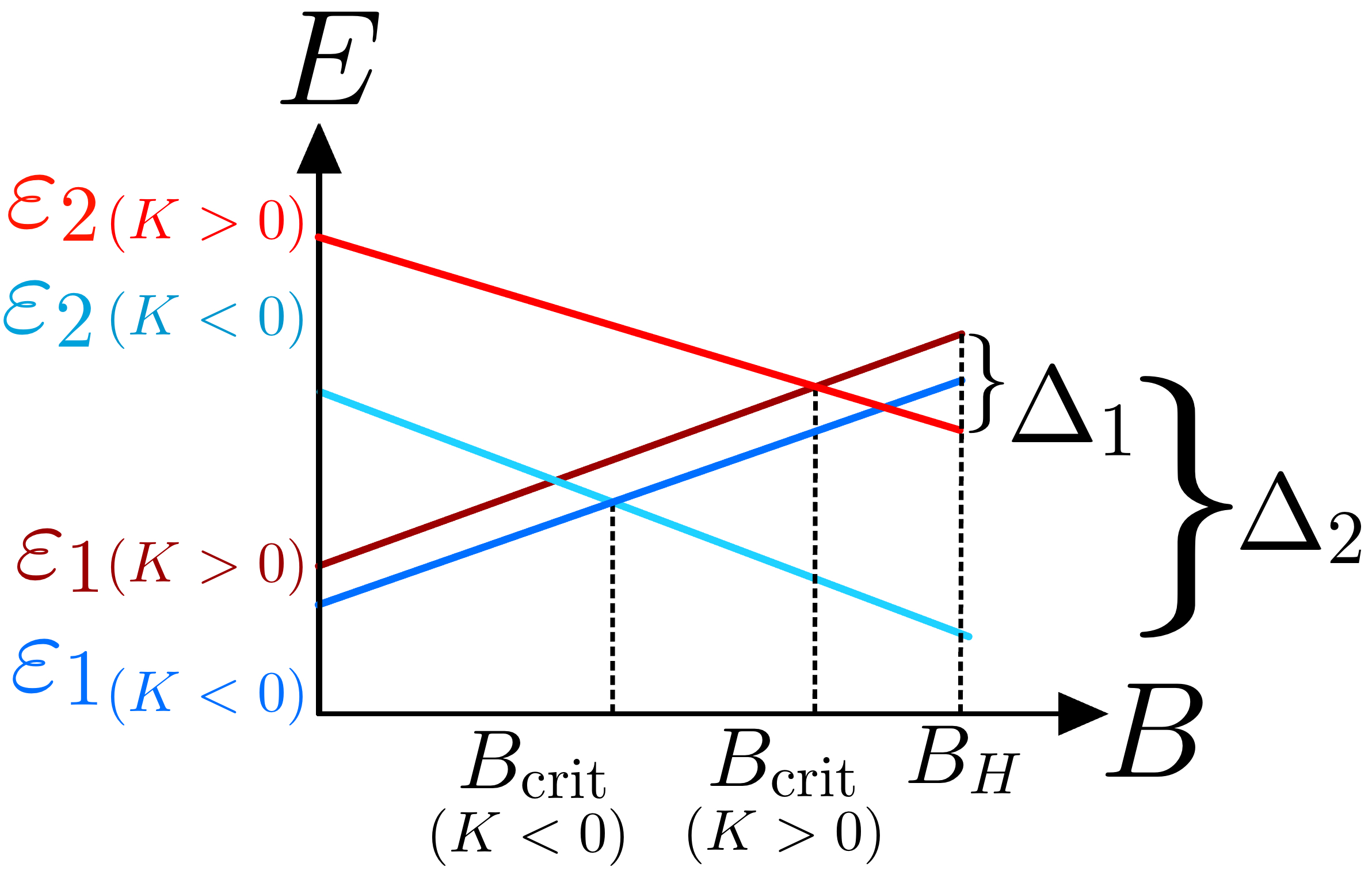}
    \caption{ { The ground state $\varepsilon_1$ and the first excited state $\varepsilon_2$ are plotted against the external magnetic field $B$ for different anisotropy values $K$. The crossing point is the Quantum Critical Point at $B_{\text{crit}}$ for both positive and negative anisotropies. $\Delta_1$ and $\Delta_2$ represent the separation of energies for positive and negative anisotropies, respectively, at a fixed high magnetic field $B_H$. }}
    \label{appendice_2_fig_2}
\end{figure}
{It is important to note that under these approximations on the cycle, the total work is non-zero and is proportional to the temperature difference between the thermal reservoirs and to the natural logarithm of $g_{1}$:}
 
\begin{equation}
    W\approx \left(T_{H}- T_{L}\right) \ln\left(g_1\right).
\end{equation}
 
{Finally, it is important to point out that the role played by anisotropy in our case is to modify the point where the QCP occurs and consequently changes the value of the energy difference $\Delta$ at the starting point of the cycle at the high magnetic field, $B=B_{H}$. We can see from Fig.~\ref{appendice_2_fig_2} that the negative anisotropy ($K<0$) improves the energy differences at a fixed $B_{H}$; on the contrary, when $K > 0$, a considerable decrease of the energy differences for the same $B_{H}$ is observed, favoring the occupation of both energy levels. This means that for $K > 0$, we do not find the Carnot efficiency for the temperatures used, and the approximations described in this appendix are not applicable.}
}


\begin{thebibliography}{47}%
\makeatletter
\providecommand \@ifxundefined [1]{%
 \@ifx{#1\undefined}
}%
\providecommand \@ifnum [1]{%
 \ifnum #1\expandafter \@firstoftwo
 \else \expandafter \@secondoftwo
 \fi
}%
\providecommand \@ifx [1]{%
 \ifx #1\expandafter \@firstoftwo
 \else \expandafter \@secondoftwo
 \fi
}%
\providecommand \natexlab [1]{#1}%
\providecommand \enquote  [1]{``#1''}%
\providecommand \bibnamefont  [1]{#1}%
\providecommand \bibfnamefont [1]{#1}%
\providecommand \citenamefont [1]{#1}%
\providecommand \href@noop [0]{\@secondoftwo}%
\providecommand \href [0]{\begingroup \@sanitize@url \@href}%
\providecommand \@href[1]{\@@startlink{#1}\@@href}%
\providecommand \@@href[1]{\endgroup#1\@@endlink}%
\providecommand \@sanitize@url [0]{\catcode `\\12\catcode `\$12\catcode `\&12\catcode `\#12\catcode `\^12\catcode `\_12\catcode `\%12\relax}%
\providecommand \@@startlink[1]{}%
\providecommand \@@endlink[0]{}%
\providecommand \url  [0]{\begingroup\@sanitize@url \@url }%
\providecommand \@url [1]{\endgroup\@href {#1}{\urlprefix }}%
\providecommand \urlprefix  [0]{URL }%
\providecommand \Eprint [0]{\href }%
\providecommand \doibase [0]{https://doi.org/}%
\providecommand \selectlanguage [0]{\@gobble}%
\providecommand \bibinfo  [0]{\@secondoftwo}%
\providecommand \bibfield  [0]{\@secondoftwo}%
\providecommand \translation [1]{[#1]}%
\providecommand \BibitemOpen [0]{}%
\providecommand \bibitemStop [0]{}%
\providecommand \bibitemNoStop [0]{.\EOS\space}%
\providecommand \EOS [0]{\spacefactor3000\relax}%
\providecommand \BibitemShut  [1]{\csname bibitem#1\endcsname}%
\let\auto@bib@innerbib\@empty
\bibitem [{\citenamefont {Geva}\ and\ \citenamefont {Kosloff}(1992{\natexlab{a}})}]{WOS:A1992HE74300067}%
  \BibitemOpen
  \bibfield  {author} {\bibinfo {author} {\bibfnamefont {E.}~\bibnamefont {Geva}}\ and\ \bibinfo {author} {\bibfnamefont {R.}~\bibnamefont {Kosloff}},\ }\bibfield  {title} {\bibinfo {title} {A quantum-mechanical heat engine operating in finite-time, a model consisting of spin-1/2 systems as the working fluid},\ }\href {https://doi.org/10.1063/1.461951} {\bibfield  {journal} {\bibinfo  {journal} {JOURNAL OF CHEMICAL PHYSICS}\ }\textbf {\bibinfo {volume} {96}},\ \bibinfo {pages} {3054} (\bibinfo {year} {1992}{\natexlab{a}})}\BibitemShut {NoStop}%
\bibitem [{\citenamefont {Geva}\ and\ \citenamefont {Kosloff}(1992{\natexlab{b}})}]{WOS:A1992JN14600053}%
  \BibitemOpen
  \bibfield  {author} {\bibinfo {author} {\bibfnamefont {E.}~\bibnamefont {Geva}}\ and\ \bibinfo {author} {\bibfnamefont {R.}~\bibnamefont {Kosloff}},\ }\bibfield  {title} {\bibinfo {title} {On the classical limit of quantum thermodynamics in finite-time},\ }\href {https://doi.org/10.1063/1.463909} {\bibfield  {journal} {\bibinfo  {journal} {JOURNAL OF CHEMICAL PHYSICS}\ }\textbf {\bibinfo {volume} {97}},\ \bibinfo {pages} {4398} (\bibinfo {year} {1992}{\natexlab{b}})}\BibitemShut {NoStop}%
\bibitem [{\citenamefont {He}\ and\ \citenamefont {Chen}(2002)}]{WOS:000174548900068}%
  \BibitemOpen
  \bibfield  {author} {\bibinfo {author} {\bibfnamefont {J.}~\bibnamefont {He}}\ and\ \bibinfo {author} {\bibfnamefont {J.}~\bibnamefont {Chen}},\ }\bibfield  {title} {\bibinfo {title} {Quantum refrigeration cycles using spin-(1)/(2) systems as the working substance},\ }\bibfield  {journal} {\bibinfo  {journal} {PHYSICAL REVIEW E}\ }\textbf {\bibinfo {volume} {65}},\ \href {https://doi.org/10.1103/PhysRevE.65.036145} {10.1103/PhysRevE.65.036145} (\bibinfo {year} {2002})\BibitemShut {NoStop}%
\bibitem [{\citenamefont {Zhang}\ \emph {et~al.}(2007)\citenamefont {Zhang}, \citenamefont {Liu}, \citenamefont {Chen},\ and\ \citenamefont {Li}}]{WOS:000247624300019}%
  \BibitemOpen
  \bibfield  {author} {\bibinfo {author} {\bibfnamefont {T.}~\bibnamefont {Zhang}}, \bibinfo {author} {\bibfnamefont {W.-T.}\ \bibnamefont {Liu}}, \bibinfo {author} {\bibfnamefont {P.-X.}\ \bibnamefont {Chen}},\ and\ \bibinfo {author} {\bibfnamefont {C.-Z.}\ \bibnamefont {Li}},\ }\bibfield  {title} {\bibinfo {title} {Four-level entangled quantum heat engines},\ }\bibfield  {journal} {\bibinfo  {journal} {PHYSICAL REVIEW A}\ }\textbf {\bibinfo {volume} {75}},\ \href {https://doi.org/10.1103/PhysRevA.75.062102} {10.1103/PhysRevA.75.062102} (\bibinfo {year} {2007})\BibitemShut {NoStop}%
\bibitem [{\citenamefont {Henrich}\ \emph {et~al.}(2007{\natexlab{a}})\citenamefont {Henrich}, \citenamefont {Mahler},\ and\ \citenamefont {Michel}}]{WOS:000246890100029}%
  \BibitemOpen
  \bibfield  {author} {\bibinfo {author} {\bibfnamefont {M.~J.}\ \bibnamefont {Henrich}}, \bibinfo {author} {\bibfnamefont {G.}~\bibnamefont {Mahler}},\ and\ \bibinfo {author} {\bibfnamefont {M.}~\bibnamefont {Michel}},\ }\bibfield  {title} {\bibinfo {title} {Driven spin systems as quantum thermodynamic machines: Fundamental limits},\ }\bibfield  {journal} {\bibinfo  {journal} {PHYSICAL REVIEW E}\ }\textbf {\bibinfo {volume} {75}},\ \href {https://doi.org/10.1103/PhysRevE.75.051118} {10.1103/PhysRevE.75.051118} (\bibinfo {year} {2007}{\natexlab{a}})\BibitemShut {NoStop}%
\bibitem [{\citenamefont {Saygin}\ and\ \citenamefont {Sisman}(2001)}]{WOS:000170647500068}%
  \BibitemOpen
  \bibfield  {author} {\bibinfo {author} {\bibfnamefont {H.}~\bibnamefont {Saygin}}\ and\ \bibinfo {author} {\bibfnamefont {A.}~\bibnamefont {Sisman}},\ }\bibfield  {title} {\bibinfo {title} {Quantum degeneracy effect on the work output from a stirling cycle},\ }\href {https://doi.org/10.1063/1.1396831} {\bibfield  {journal} {\bibinfo  {journal} {JOURNAL OF APPLIED PHYSICS}\ }\textbf {\bibinfo {volume} {90}},\ \bibinfo {pages} {3086} (\bibinfo {year} {2001})}\BibitemShut {NoStop}%
\bibitem [{\citenamefont {Zhang}(2008)}]{WOS:000259449100016}%
  \BibitemOpen
  \bibfield  {author} {\bibinfo {author} {\bibfnamefont {G.~F.}\ \bibnamefont {Zhang}},\ }\bibfield  {title} {\bibinfo {title} {Entangled quantum heat engines based on two two-spin systems with dzyaloshinski-moriya anisotropic antisymmetric interaction},\ }\href {https://doi.org/10.1140/epjd/e2008-00133-0} {\bibfield  {journal} {\bibinfo  {journal} {EUROPEAN PHYSICAL JOURNAL D}\ }\textbf {\bibinfo {volume} {49}},\ \bibinfo {pages} {123} (\bibinfo {year} {2008})}\BibitemShut {NoStop}%
\bibitem [{\citenamefont {Cakmak}\ and\ \citenamefont {Mustecaplioglu}(2019)}]{WOS:000460663800003}%
  \BibitemOpen
  \bibfield  {author} {\bibinfo {author} {\bibfnamefont {B.}~\bibnamefont {Cakmak}}\ and\ \bibinfo {author} {\bibfnamefont {O.~E.}\ \bibnamefont {Mustecaplioglu}},\ }\bibfield  {title} {\bibinfo {title} {Spin quantum heat engines with shortcuts to adiabaticity},\ }\bibfield  {journal} {\bibinfo  {journal} {PHYSICAL REVIEW E}\ }\textbf {\bibinfo {volume} {99}},\ \href {https://doi.org/10.1103/PhysRevE.99.032108} {10.1103/PhysRevE.99.032108} (\bibinfo {year} {2019})\BibitemShut {NoStop}%
\bibitem [{\citenamefont {Wu}\ \emph {et~al.}(2006)\citenamefont {Wu}, \citenamefont {Chen}, \citenamefont {Wu}, \citenamefont {Sun},\ and\ \citenamefont {Wu}}]{WOS:000238758700035}%
  \BibitemOpen
  \bibfield  {author} {\bibinfo {author} {\bibfnamefont {F.}~\bibnamefont {Wu}}, \bibinfo {author} {\bibfnamefont {L.}~\bibnamefont {Chen}}, \bibinfo {author} {\bibfnamefont {S.}~\bibnamefont {Wu}}, \bibinfo {author} {\bibfnamefont {F.}~\bibnamefont {Sun}},\ and\ \bibinfo {author} {\bibfnamefont {C.}~\bibnamefont {Wu}},\ }\bibfield  {title} {\bibinfo {title} {Performance of an irreversible quantum carnot engine with spin 1/2},\ }\bibfield  {journal} {\bibinfo  {journal} {JOURNAL OF CHEMICAL PHYSICS}\ }\textbf {\bibinfo {volume} {124}},\ \href {https://doi.org/10.1063/1.2200693} {10.1063/1.2200693} (\bibinfo {year} {2006})\BibitemShut {NoStop}%
\bibitem [{\citenamefont {Azimi}\ \emph {et~al.}(2014)\citenamefont {Azimi}, \citenamefont {Chotorlishvili}, \citenamefont {Mishra}, \citenamefont {Vekua}, \citenamefont {Huebner},\ and\ \citenamefont {Berakdar}}]{WOS:000339078100001}%
  \BibitemOpen
  \bibfield  {author} {\bibinfo {author} {\bibfnamefont {M.}~\bibnamefont {Azimi}}, \bibinfo {author} {\bibfnamefont {L.}~\bibnamefont {Chotorlishvili}}, \bibinfo {author} {\bibfnamefont {S.~K.}\ \bibnamefont {Mishra}}, \bibinfo {author} {\bibfnamefont {T.}~\bibnamefont {Vekua}}, \bibinfo {author} {\bibfnamefont {W.}~\bibnamefont {Huebner}},\ and\ \bibinfo {author} {\bibfnamefont {J.}~\bibnamefont {Berakdar}},\ }\bibfield  {title} {\bibinfo {title} {Quantum otto heat engine based on a multiferroic chain working substance},\ }\bibfield  {journal} {\bibinfo  {journal} {NEW JOURNAL OF PHYSICS}\ }\textbf {\bibinfo {volume} {16}},\ \href {https://doi.org/10.1088/1367-2630/16/6/063018} {10.1088/1367-2630/16/6/063018} (\bibinfo {year} {2014})\BibitemShut {NoStop}%
\bibitem [{\citenamefont {Allahverdyan}\ \emph {et~al.}(2005)\citenamefont {Allahverdyan}, \citenamefont {Gracia},\ and\ \citenamefont {Nieuwenhuizen}}]{WOS:000228752500014}%
  \BibitemOpen
  \bibfield  {author} {\bibinfo {author} {\bibfnamefont {A.}~\bibnamefont {Allahverdyan}}, \bibinfo {author} {\bibfnamefont {R.}~\bibnamefont {Gracia}},\ and\ \bibinfo {author} {\bibfnamefont {T.}~\bibnamefont {Nieuwenhuizen}},\ }\bibfield  {title} {\bibinfo {title} {Work extraction in the spin-boson model},\ }\bibfield  {journal} {\bibinfo  {journal} {PHYSICAL REVIEW E}\ }\textbf {\bibinfo {volume} {71}},\ \href {https://doi.org/10.1103/PhysRevE.71.046106} {10.1103/PhysRevE.71.046106} (\bibinfo {year} {2005})\BibitemShut {NoStop}%
\bibitem [{\citenamefont {Henrich}\ \emph {et~al.}(2007{\natexlab{b}})\citenamefont {Henrich}, \citenamefont {Rempp},\ and\ \citenamefont {Mahler}}]{WOS:000251858800016}%
  \BibitemOpen
  \bibfield  {author} {\bibinfo {author} {\bibfnamefont {M.~J.}\ \bibnamefont {Henrich}}, \bibinfo {author} {\bibfnamefont {F.}~\bibnamefont {Rempp}},\ and\ \bibinfo {author} {\bibfnamefont {G.}~\bibnamefont {Mahler}},\ }\bibfield  {title} {\bibinfo {title} {Quantum thermodynamic otto machines: A spin-system approach},\ }\href {https://doi.org/10.1140/epjst/e2007-00371-8} {\bibfield  {journal} {\bibinfo  {journal} {EUROPEAN PHYSICAL JOURNAL-SPECIAL TOPICS}\ }\textbf {\bibinfo {volume} {151}},\ \bibinfo {pages} {157} (\bibinfo {year} {2007}{\natexlab{b}})}\BibitemShut {NoStop}%
\bibitem [{\citenamefont {Wang}\ \emph {et~al.}(2007)\citenamefont {Wang}, \citenamefont {He},\ and\ \citenamefont {Xin}}]{WOS:000243928200018}%
  \BibitemOpen
  \bibfield  {author} {\bibinfo {author} {\bibfnamefont {J.}~\bibnamefont {Wang}}, \bibinfo {author} {\bibfnamefont {J.}~\bibnamefont {He}},\ and\ \bibinfo {author} {\bibfnamefont {Y.}~\bibnamefont {Xin}},\ }\bibfield  {title} {\bibinfo {title} {Performance analysis of a spin quantum heat engine cycle with internal friction},\ }\href {https://doi.org/10.1088/0031-8949/75/2/018} {\bibfield  {journal} {\bibinfo  {journal} {PHYSICA SCRIPTA}\ }\textbf {\bibinfo {volume} {75}},\ \bibinfo {pages} {227} (\bibinfo {year} {2007})}\BibitemShut {NoStop}%
\bibitem [{\citenamefont {Chen}\ \emph {et~al.}(2002)\citenamefont {Chen}, \citenamefont {Lin},\ and\ \citenamefont {Hua}}]{WOS:000178091500025}%
  \BibitemOpen
  \bibfield  {author} {\bibinfo {author} {\bibfnamefont {J.}~\bibnamefont {Chen}}, \bibinfo {author} {\bibfnamefont {B.}~\bibnamefont {Lin}},\ and\ \bibinfo {author} {\bibfnamefont {B.}~\bibnamefont {Hua}},\ }\bibfield  {title} {\bibinfo {title} {The performance of a quantum heat engine working with spin systems},\ }\href {https://doi.org/10.1088/0022-3727/35/16/322} {\bibfield  {journal} {\bibinfo  {journal} {JOURNAL OF PHYSICS D-APPLIED PHYSICS}\ }\textbf {\bibinfo {volume} {35}},\ \bibinfo {pages} {2051} (\bibinfo {year} {2002})}\BibitemShut {NoStop}%
\bibitem [{\citenamefont {Ono}\ \emph {et~al.}(2020)\citenamefont {Ono}, \citenamefont {Shevchenko}, \citenamefont {Mori}, \citenamefont {Moriyama},\ and\ \citenamefont {Nori}}]{WOS:000577236900006}%
  \BibitemOpen
  \bibfield  {author} {\bibinfo {author} {\bibfnamefont {K.}~\bibnamefont {Ono}}, \bibinfo {author} {\bibfnamefont {S.~N.}\ \bibnamefont {Shevchenko}}, \bibinfo {author} {\bibfnamefont {T.}~\bibnamefont {Mori}}, \bibinfo {author} {\bibfnamefont {S.}~\bibnamefont {Moriyama}},\ and\ \bibinfo {author} {\bibfnamefont {F.}~\bibnamefont {Nori}},\ }\bibfield  {title} {\bibinfo {title} {Analog of a quantum heat engine using a single-spin qubit},\ }\bibfield  {journal} {\bibinfo  {journal} {PHYSICAL REVIEW LETTERS}\ }\textbf {\bibinfo {volume} {125}},\ \href {https://doi.org/10.1103/PhysRevLett.125.166802} {10.1103/PhysRevLett.125.166802} (\bibinfo {year} {2020})\BibitemShut {NoStop}%
\bibitem [{\citenamefont {Altintas}\ and\ \citenamefont {Mustecaplioglu}(2015)}]{WOS:000360065300001}%
  \BibitemOpen
  \bibfield  {author} {\bibinfo {author} {\bibfnamefont {F.}~\bibnamefont {Altintas}}\ and\ \bibinfo {author} {\bibfnamefont {O.~E.}\ \bibnamefont {Mustecaplioglu}},\ }\bibfield  {title} {\bibinfo {title} {General formalism of local thermodynamics with an example: Quantum otto engine with a spin-1/2 coupled to an arbitrary spin},\ }\bibfield  {journal} {\bibinfo  {journal} {PHYSICAL REVIEW E}\ }\textbf {\bibinfo {volume} {92}},\ \href {https://doi.org/10.1103/PhysRevE.92.022142} {10.1103/PhysRevE.92.022142} (\bibinfo {year} {2015})\BibitemShut {NoStop}%
\bibitem [{\citenamefont {Wu}\ \emph {et~al.}(1998)\citenamefont {Wu}, \citenamefont {Chen}, \citenamefont {Sun}, \citenamefont {Wu},\ and\ \citenamefont {Hua}}]{WOS:000074107900005}%
  \BibitemOpen
  \bibfield  {author} {\bibinfo {author} {\bibfnamefont {F.}~\bibnamefont {Wu}}, \bibinfo {author} {\bibfnamefont {L.}~\bibnamefont {Chen}}, \bibinfo {author} {\bibfnamefont {F.}~\bibnamefont {Sun}}, \bibinfo {author} {\bibfnamefont {C.}~\bibnamefont {Wu}},\ and\ \bibinfo {author} {\bibfnamefont {P.}~\bibnamefont {Hua}},\ }\bibfield  {title} {\bibinfo {title} {Optimum performance parameters for a quantum carnot heat pump with spin-1/2},\ }\href {https://doi.org/10.1016/S0196-8904(98)00004-1} {\bibfield  {journal} {\bibinfo  {journal} {ENERGY CONVERSION AND MANAGEMENT}\ }\textbf {\bibinfo {volume} {39}},\ \bibinfo {pages} {1161} (\bibinfo {year} {1998})}\BibitemShut {NoStop}%
\bibitem [{\citenamefont {Alecce}\ \emph {et~al.}(2015)\citenamefont {Alecce}, \citenamefont {Galve}, \citenamefont {Lo~Gullo}, \citenamefont {Dell'Anna}, \citenamefont {Plastina},\ and\ \citenamefont {Zambrini}}]{WOS:000359128100004}%
  \BibitemOpen
  \bibfield  {author} {\bibinfo {author} {\bibfnamefont {A.}~\bibnamefont {Alecce}}, \bibinfo {author} {\bibfnamefont {F.}~\bibnamefont {Galve}}, \bibinfo {author} {\bibfnamefont {N.}~\bibnamefont {Lo~Gullo}}, \bibinfo {author} {\bibfnamefont {L.}~\bibnamefont {Dell'Anna}}, \bibinfo {author} {\bibfnamefont {F.}~\bibnamefont {Plastina}},\ and\ \bibinfo {author} {\bibfnamefont {R.}~\bibnamefont {Zambrini}},\ }\bibfield  {title} {\bibinfo {title} {Quantum otto cycle with inner friction: finite-time and disorder effects},\ }\bibfield  {journal} {\bibinfo  {journal} {NEW JOURNAL OF PHYSICS}\ }\textbf {\bibinfo {volume} {17}},\ \href {https://doi.org/10.1088/1367-2630/17/7/075007} {10.1088/1367-2630/17/7/075007} (\bibinfo {year} {2015})\BibitemShut {NoStop}%
\bibitem [{\citenamefont {Kosloff}\ and\ \citenamefont {Feldmann}(2010)}]{WOS:000280233300002}%
  \BibitemOpen
  \bibfield  {author} {\bibinfo {author} {\bibfnamefont {R.}~\bibnamefont {Kosloff}}\ and\ \bibinfo {author} {\bibfnamefont {T.}~\bibnamefont {Feldmann}},\ }\bibfield  {title} {\bibinfo {title} {Optimal performance of reciprocating demagnetization quantum refrigerators},\ }\bibfield  {journal} {\bibinfo  {journal} {PHYSICAL REVIEW E}\ }\textbf {\bibinfo {volume} {82}},\ \href {https://doi.org/10.1103/PhysRevE.82.011134} {10.1103/PhysRevE.82.011134} (\bibinfo {year} {2010})\BibitemShut {NoStop}%
\bibitem [{\citenamefont {Katz}\ and\ \citenamefont {Kosloff}(2016)}]{WOS:000377262900029}%
  \BibitemOpen
  \bibfield  {author} {\bibinfo {author} {\bibfnamefont {G.}~\bibnamefont {Katz}}\ and\ \bibinfo {author} {\bibfnamefont {R.}~\bibnamefont {Kosloff}},\ }\bibfield  {title} {\bibinfo {title} {Quantum thermodynamics in strong coupling: Heat transport and refrigeration},\ }\bibfield  {journal} {\bibinfo  {journal} {ENTROPY}\ }\textbf {\bibinfo {volume} {18}},\ \href {https://doi.org/10.3390/e18050186} {10.3390/e18050186} (\bibinfo {year} {2016})\BibitemShut {NoStop}%
\bibitem [{\citenamefont {Altintas}\ \emph {et~al.}(2014)\citenamefont {Altintas}, \citenamefont {Hardal},\ and\ \citenamefont {Mustecaplioglu}}]{WOS:000341247200002}%
  \BibitemOpen
  \bibfield  {author} {\bibinfo {author} {\bibfnamefont {F.}~\bibnamefont {Altintas}}, \bibinfo {author} {\bibfnamefont {A.~U.~C.}\ \bibnamefont {Hardal}},\ and\ \bibinfo {author} {\bibfnamefont {O.~E.}\ \bibnamefont {Mustecaplioglu}},\ }\bibfield  {title} {\bibinfo {title} {Quantum correlated heat engine with spin squeezing},\ }\bibfield  {journal} {\bibinfo  {journal} {PHYSICAL REVIEW E}\ }\textbf {\bibinfo {volume} {90}},\ \href {https://doi.org/10.1103/PhysRevE.90.032102} {10.1103/PhysRevE.90.032102} (\bibinfo {year} {2014})\BibitemShut {NoStop}%
\bibitem [{\citenamefont {D\"ur}\ \emph {et~al.}(2000)\citenamefont {D\"ur}, \citenamefont {Vidal},\ and\ \citenamefont {Cirac}}]{PhysRevA.62.062314}%
  \BibitemOpen
  \bibfield  {author} {\bibinfo {author} {\bibfnamefont {W.}~\bibnamefont {D\"ur}}, \bibinfo {author} {\bibfnamefont {G.}~\bibnamefont {Vidal}},\ and\ \bibinfo {author} {\bibfnamefont {J.~I.}\ \bibnamefont {Cirac}},\ }\bibfield  {title} {\bibinfo {title} {Three qubits can be entangled in two inequivalent ways},\ }\href {https://doi.org/10.1103/PhysRevA.62.062314} {\bibfield  {journal} {\bibinfo  {journal} {Phys. Rev. A}\ }\textbf {\bibinfo {volume} {62}},\ \bibinfo {pages} {062314} (\bibinfo {year} {2000})}\BibitemShut {NoStop}%
\bibitem [{\citenamefont {Kosloff}\ and\ \citenamefont {Rezek}(2017)}]{WOS:000400579500001}%
  \BibitemOpen
  \bibfield  {author} {\bibinfo {author} {\bibfnamefont {R.}~\bibnamefont {Kosloff}}\ and\ \bibinfo {author} {\bibfnamefont {Y.}~\bibnamefont {Rezek}},\ }\bibfield  {title} {\bibinfo {title} {The quantum harmonic otto cycle},\ }\bibfield  {journal} {\bibinfo  {journal} {ENTROPY}\ }\textbf {\bibinfo {volume} {19}},\ \href {https://doi.org/10.3390/e19040136} {10.3390/e19040136} (\bibinfo {year} {2017})\BibitemShut {NoStop}%
\bibitem [{\citenamefont {Dann}\ \emph {et~al.}(2020)\citenamefont {Dann}, \citenamefont {Kosloff},\ and\ \citenamefont {Salamon}}]{WOS:000592899500001}%
  \BibitemOpen
  \bibfield  {author} {\bibinfo {author} {\bibfnamefont {R.}~\bibnamefont {Dann}}, \bibinfo {author} {\bibfnamefont {R.}~\bibnamefont {Kosloff}},\ and\ \bibinfo {author} {\bibfnamefont {P.}~\bibnamefont {Salamon}},\ }\bibfield  {title} {\bibinfo {title} {Quantum finite-time thermodynamics: Insight from a single qubit engine},\ }\bibfield  {journal} {\bibinfo  {journal} {ENTROPY}\ }\textbf {\bibinfo {volume} {22}},\ \href {https://doi.org/10.3390/e22111255} {10.3390/e22111255} (\bibinfo {year} {2020})\BibitemShut {NoStop}%
\bibitem [{\citenamefont {Deffner}(2018)}]{WOS:000451308800067}%
  \BibitemOpen
  \bibfield  {author} {\bibinfo {author} {\bibfnamefont {S.}~\bibnamefont {Deffner}},\ }\bibfield  {title} {\bibinfo {title} {Efficiency of harmonic quantum otto engines at maximal power},\ }\bibfield  {journal} {\bibinfo  {journal} {ENTROPY}\ }\textbf {\bibinfo {volume} {20}},\ \href {https://doi.org/10.3390/e20110875} {10.3390/e20110875} (\bibinfo {year} {2018})\BibitemShut {NoStop}%
\bibitem [{\citenamefont {Smith}\ \emph {et~al.}(2020)\citenamefont {Smith}, \citenamefont {Pal},\ and\ \citenamefont {Deffner}}]{WOS:000544256400007}%
  \BibitemOpen
  \bibfield  {author} {\bibinfo {author} {\bibfnamefont {Z.}~\bibnamefont {Smith}}, \bibinfo {author} {\bibfnamefont {P.~S.}\ \bibnamefont {Pal}},\ and\ \bibinfo {author} {\bibfnamefont {S.}~\bibnamefont {Deffner}},\ }\bibfield  {title} {\bibinfo {title} {Endoreversible otto engines at maximal power},\ }\href {https://doi.org/10.1515/jnet-2020-0039} {\bibfield  {journal} {\bibinfo  {journal} {JOURNAL OF NON-EQUILIBRIUM THERMODYNAMICS}\ }\textbf {\bibinfo {volume} {45}},\ \bibinfo {pages} {305} (\bibinfo {year} {2020})}\BibitemShut {NoStop}%
\bibitem [{\citenamefont {Reid}\ \emph {et~al.}(2017)\citenamefont {Reid}, \citenamefont {Pigeon}, \citenamefont {Antezza},\ and\ \citenamefont {De~Chiara}}]{WOS:000426262900001}%
  \BibitemOpen
  \bibfield  {author} {\bibinfo {author} {\bibfnamefont {B.}~\bibnamefont {Reid}}, \bibinfo {author} {\bibfnamefont {S.}~\bibnamefont {Pigeon}}, \bibinfo {author} {\bibfnamefont {M.}~\bibnamefont {Antezza}},\ and\ \bibinfo {author} {\bibfnamefont {G.}~\bibnamefont {De~Chiara}},\ }\bibfield  {title} {\bibinfo {title} {A self-contained quantum harmonic engine},\ }\bibfield  {journal} {\bibinfo  {journal} {EPL}\ }\textbf {\bibinfo {volume} {120}},\ \href {https://doi.org/10.1209/0295-5075/120/60006} {10.1209/0295-5075/120/60006} (\bibinfo {year} {2017})\BibitemShut {NoStop}%
\bibitem [{\citenamefont {Hewgill}\ \emph {et~al.}(2018)\citenamefont {Hewgill}, \citenamefont {Ferraro},\ and\ \citenamefont {De~Chiara}}]{WOS:000446164500002}%
  \BibitemOpen
  \bibfield  {author} {\bibinfo {author} {\bibfnamefont {A.}~\bibnamefont {Hewgill}}, \bibinfo {author} {\bibfnamefont {A.}~\bibnamefont {Ferraro}},\ and\ \bibinfo {author} {\bibfnamefont {G.}~\bibnamefont {De~Chiara}},\ }\bibfield  {title} {\bibinfo {title} {Quantum correlations and thermodynamic performances of two-qubit engines with local and common baths},\ }\bibfield  {journal} {\bibinfo  {journal} {PHYSICAL REVIEW A}\ }\textbf {\bibinfo {volume} {98}},\ \href {https://doi.org/10.1103/PhysRevA.98.042102} {10.1103/PhysRevA.98.042102} (\bibinfo {year} {2018})\BibitemShut {NoStop}%
\bibitem [{\citenamefont {Pena}\ \emph {et~al.}(2020)\citenamefont {Pena}, \citenamefont {Zambrano}, \citenamefont {Negrete}, \citenamefont {De~Chiara}, \citenamefont {Orellana},\ and\ \citenamefont {Vargas}}]{WOS:000506846500003}%
  \BibitemOpen
  \bibfield  {author} {\bibinfo {author} {\bibfnamefont {F.~J.}\ \bibnamefont {Pena}}, \bibinfo {author} {\bibfnamefont {D.}~\bibnamefont {Zambrano}}, \bibinfo {author} {\bibfnamefont {O.}~\bibnamefont {Negrete}}, \bibinfo {author} {\bibfnamefont {G.}~\bibnamefont {De~Chiara}}, \bibinfo {author} {\bibfnamefont {P.~A.}\ \bibnamefont {Orellana}},\ and\ \bibinfo {author} {\bibfnamefont {P.}~\bibnamefont {Vargas}},\ }\bibfield  {title} {\bibinfo {title} {Quasistatic and quantum-adiabatic otto engine for a two-dimensional material: The case of a graphene quantum dot},\ }\bibfield  {journal} {\bibinfo  {journal} {PHYSICAL REVIEW E}\ }\textbf {\bibinfo {volume} {101}},\ \href {https://doi.org/10.1103/PhysRevE.101.012116} {10.1103/PhysRevE.101.012116} (\bibinfo {year} {2020})\BibitemShut {NoStop}%
\bibitem [{\citenamefont {Hamedani~Raja}\ \emph {et~al.}(2021)\citenamefont {Hamedani~Raja}, \citenamefont {Maniscalco}, \citenamefont {Paraoanu}, \citenamefont {Pekola},\ and\ \citenamefont {Lo~Gullo}}]{WOS:000630746700001}%
  \BibitemOpen
  \bibfield  {author} {\bibinfo {author} {\bibfnamefont {S.}~\bibnamefont {Hamedani~Raja}}, \bibinfo {author} {\bibfnamefont {S.}~\bibnamefont {Maniscalco}}, \bibinfo {author} {\bibfnamefont {G.~S.}\ \bibnamefont {Paraoanu}}, \bibinfo {author} {\bibfnamefont {J.~P.}\ \bibnamefont {Pekola}},\ and\ \bibinfo {author} {\bibfnamefont {N.}~\bibnamefont {Lo~Gullo}},\ }\bibfield  {title} {\bibinfo {title} {Finite-time quantum stirling heat engine},\ }\bibfield  {journal} {\bibinfo  {journal} {NEW JOURNAL OF PHYSICS}\ }\textbf {\bibinfo {volume} {23}},\ \href {https://doi.org/10.1088/1367-2630/abe9d7} {10.1088/1367-2630/abe9d7} (\bibinfo {year} {2021})\BibitemShut {NoStop}%
\bibitem [{\citenamefont {Pili}\ \emph {et~al.}(2023)\citenamefont {Pili}, \citenamefont {Khordad}, \citenamefont {Sedehi},\ and\ \citenamefont {Avazpour}}]{WOS:001054373800001}%
  \BibitemOpen
  \bibfield  {author} {\bibinfo {author} {\bibfnamefont {A.~H.~B.}\ \bibnamefont {Pili}}, \bibinfo {author} {\bibfnamefont {R.}~\bibnamefont {Khordad}}, \bibinfo {author} {\bibfnamefont {H.~R.~R.}\ \bibnamefont {Sedehi}},\ and\ \bibinfo {author} {\bibfnamefont {A.}~\bibnamefont {Avazpour}},\ }\bibfield  {title} {\bibinfo {title} {Study of performance of quantum stirling engine using 2d and 3d heisenberg model},\ }\bibfield  {journal} {\bibinfo  {journal} {INTERNATIONAL JOURNAL OF THEORETICAL PHYSICS}\ }\textbf {\bibinfo {volume} {62}},\ \href {https://doi.org/10.1007/s10773-023-05450-5} {10.1007/s10773-023-05450-5} (\bibinfo {year} {2023})\BibitemShut {NoStop}%
\bibitem [{\citenamefont {Yin}\ \emph {et~al.}(2018)\citenamefont {Yin}, \citenamefont {Chen},\ and\ \citenamefont {Wu}}]{WOS:000452093900005}%
  \BibitemOpen
  \bibfield  {author} {\bibinfo {author} {\bibfnamefont {Y.}~\bibnamefont {Yin}}, \bibinfo {author} {\bibfnamefont {L.}~\bibnamefont {Chen}},\ and\ \bibinfo {author} {\bibfnamefont {F.}~\bibnamefont {Wu}},\ }\bibfield  {title} {\bibinfo {title} {Performance of quantum stirling heat engine with numerous copies of extreme relativistic particles confined in 1d potential well},\ }\href {https://doi.org/10.1016/j.physa.2018.02.202} {\bibfield  {journal} {\bibinfo  {journal} {PHYSICA A-STATISTICAL MECHANICS AND ITS APPLICATIONS}\ }\textbf {\bibinfo {volume} {503}},\ \bibinfo {pages} {58} (\bibinfo {year} {2018})}\BibitemShut {NoStop}%
\bibitem [{\citenamefont {Peña}\ \emph {et~al.}(2020)\citenamefont {Peña}, \citenamefont {Negrete}, \citenamefont {Cortés},\ and\ \citenamefont {Vargas}}]{pena1}%
  \BibitemOpen
  \bibfield  {author} {\bibinfo {author} {\bibfnamefont {F.~J.}\ \bibnamefont {Peña}}, \bibinfo {author} {\bibfnamefont {O.}~\bibnamefont {Negrete}}, \bibinfo {author} {\bibfnamefont {N.}~\bibnamefont {Cortés}},\ and\ \bibinfo {author} {\bibfnamefont {P.}~\bibnamefont {Vargas}},\ }\bibfield  {title} {\bibinfo {title} {Otto engine: Classical and quantum approach},\ }\bibfield  {journal} {\bibinfo  {journal} {Entropy}\ }\textbf {\bibinfo {volume} {22}},\ \href {https://doi.org/10.3390/e22070755} {10.3390/e22070755} (\bibinfo {year} {2020})\BibitemShut {NoStop}%
\bibitem [{\citenamefont {Purkait}\ and\ \citenamefont {Biswas}(2022)}]{WOS:000832774500001}%
  \BibitemOpen
  \bibfield  {author} {\bibinfo {author} {\bibfnamefont {C.}~\bibnamefont {Purkait}}\ and\ \bibinfo {author} {\bibfnamefont {A.}~\bibnamefont {Biswas}},\ }\bibfield  {title} {\bibinfo {title} {Performance of heisenberg-coupled spins as quantum stirling heat machine near quantum critical point},\ }\bibfield  {journal} {\bibinfo  {journal} {PHYSICS LETTERS A}\ }\textbf {\bibinfo {volume} {442}},\ \href {https://doi.org/10.1016/j.physleta.2022.128180} {10.1016/j.physleta.2022.128180} (\bibinfo {year} {2022})\BibitemShut {NoStop}%
\bibitem [{\citenamefont {Li-Mei}\ and\ \citenamefont {Guo-Feng}(2017)}]{WOS:000425270000003}%
  \BibitemOpen
  \bibfield  {author} {\bibinfo {author} {\bibfnamefont {Z.}~\bibnamefont {Li-Mei}}\ and\ \bibinfo {author} {\bibfnamefont {Z.}~\bibnamefont {Guo-Feng}},\ }\bibfield  {title} {\bibinfo {title} {Entangled quantum otto and quantum stirling heat engine based on two-spin systems with dzyaloshinski-moriya interaction},\ }\bibfield  {journal} {\bibinfo  {journal} {ACTA PHYSICA SINICA}\ }\textbf {\bibinfo {volume} {66}},\ \href {https://doi.org/10.7498/aps.66.240502} {10.7498/aps.66.240502} (\bibinfo {year} {2017})\BibitemShut {NoStop}%
\bibitem [{\citenamefont {Cakmak}(2022)}]{WOS:000778793500034}%
  \BibitemOpen
  \bibfield  {author} {\bibinfo {author} {\bibfnamefont {S.}~\bibnamefont {Cakmak}},\ }\bibfield  {title} {\bibinfo {title} {Benchmarking quantum stirling and otto cycles for an interacting spin system},\ }\href {https://doi.org/10.1364/JOSAB.447206} {\bibfield  {journal} {\bibinfo  {journal} {JOURNAL OF THE OPTICAL SOCIETY OF AMERICA B-OPTICAL PHYSICS}\ }\textbf {\bibinfo {volume} {39}},\ \bibinfo {pages} {1209} (\bibinfo {year} {2022})}\BibitemShut {NoStop}%
\bibitem [{\citenamefont {Kuznetsova}\ \emph {et~al.}(2023)\citenamefont {Kuznetsova}, \citenamefont {Yurischev},\ and\ \citenamefont {Haddadi}}]{Kuznetsova}%
  \BibitemOpen
  \bibfield  {author} {\bibinfo {author} {\bibfnamefont {E.~I.}\ \bibnamefont {Kuznetsova}}, \bibinfo {author} {\bibfnamefont {M.~A.}\ \bibnamefont {Yurischev}},\ and\ \bibinfo {author} {\bibfnamefont {S.}~\bibnamefont {Haddadi}},\ }\bibfield  {title} {\bibinfo {title} {Quantum otto heat engines on xyz spin working medium with dm and ksea interactions: operating modes and efficiency at maximal work output},\ }\href {https://doi.org/10.1007/s11128-023-03944-z} {\bibfield  {journal} {\bibinfo  {journal} {QUANTUM INFORMATION PROCESSING}\ }\textbf {\bibinfo {volume} {22}},\ \bibinfo {pages} {192} (\bibinfo {year} {2023})}\BibitemShut {NoStop}%
\bibitem [{\citenamefont {Kamta}\ and\ \citenamefont {Starace}(2002)}]{KamtaStarace}%
  \BibitemOpen
  \bibfield  {author} {\bibinfo {author} {\bibfnamefont {G.~L.}\ \bibnamefont {Kamta}}\ and\ \bibinfo {author} {\bibfnamefont {A.~F.}\ \bibnamefont {Starace}},\ }\bibfield  {title} {\bibinfo {title} {Anisotropy and magnetic field effects on the entanglement of a two qubit heisenberg xy chain},\ }\href {https://doi.org/10.1103/PhysRevLett.88.107901} {\bibfield  {journal} {\bibinfo  {journal} {PHYSICAL REVIEW LETTERS}\ }\textbf {\bibinfo {volume} {88}},\ \bibinfo {pages} {107901} (\bibinfo {year} {2002})}\BibitemShut {NoStop}%
\bibitem [{\citenamefont {Araya}\ \emph {et~al.}(2023)\citenamefont {Araya}, \citenamefont {Pena}, \citenamefont {Norambuena}, \citenamefont {Castorene},\ and\ \citenamefont {Vargas}}]{WOS:001130894000001}%
  \BibitemOpen
  \bibfield  {author} {\bibinfo {author} {\bibfnamefont {C.}~\bibnamefont {Araya}}, \bibinfo {author} {\bibfnamefont {F.~J.}\ \bibnamefont {Pena}}, \bibinfo {author} {\bibfnamefont {A.}~\bibnamefont {Norambuena}}, \bibinfo {author} {\bibfnamefont {B.}~\bibnamefont {Castorene}},\ and\ \bibinfo {author} {\bibfnamefont {P.}~\bibnamefont {Vargas}},\ }\bibfield  {title} {\bibinfo {title} {Magnetic stirling cycle for qubits with anisotropy near the quantum critical point},\ }\bibfield  {journal} {\bibinfo  {journal} {TECHNOLOGIES}\ }\textbf {\bibinfo {volume} {11}},\ \href {https://doi.org/10.3390/technologies11060169} {10.3390/technologies11060169} (\bibinfo {year} {2023})\BibitemShut {NoStop}%
\bibitem [{\citenamefont {Xie}\ and\ \citenamefont {Eberly}(2021)}]{WOS:000677559400001}%
  \BibitemOpen
  \bibfield  {author} {\bibinfo {author} {\bibfnamefont {S.}~\bibnamefont {Xie}}\ and\ \bibinfo {author} {\bibfnamefont {J.~H.}\ \bibnamefont {Eberly}},\ }\bibfield  {title} {\bibinfo {title} {Triangle measure of tripartite entanglement},\ }\bibfield  {journal} {\bibinfo  {journal} {PHYSICAL REVIEW LETTERS}\ }\textbf {\bibinfo {volume} {127}},\ \href {https://doi.org/10.1103/PhysRevLett.127.040403} {10.1103/PhysRevLett.127.040403} (\bibinfo {year} {2021})\BibitemShut {NoStop}%
\bibitem [{\citenamefont {Takeda}\ \emph {et~al.}(2021)\citenamefont {Takeda}, \citenamefont {Noiri}, \citenamefont {Nakajima}, \citenamefont {Yoneda}, \citenamefont {Kobayashi},\ and\ \citenamefont {Tarucha}}]{WOS:000658600600002}%
  \BibitemOpen
  \bibfield  {author} {\bibinfo {author} {\bibfnamefont {K.}~\bibnamefont {Takeda}}, \bibinfo {author} {\bibfnamefont {A.}~\bibnamefont {Noiri}}, \bibinfo {author} {\bibfnamefont {T.}~\bibnamefont {Nakajima}}, \bibinfo {author} {\bibfnamefont {J.}~\bibnamefont {Yoneda}}, \bibinfo {author} {\bibfnamefont {T.}~\bibnamefont {Kobayashi}},\ and\ \bibinfo {author} {\bibfnamefont {S.}~\bibnamefont {Tarucha}},\ }\bibfield  {title} {\bibinfo {title} {Quantum tomography of an entangled three-qubit state in silicon},\ }\href {https://doi.org/10.1038/s41565-021-00925-0} {\bibfield  {journal} {\bibinfo  {journal} {NATURE NANOTECHNOLOGY}\ }\textbf {\bibinfo {volume} {16}},\ \bibinfo {pages} {965+} (\bibinfo {year} {2021})}\BibitemShut {NoStop}%
\bibitem [{\citenamefont {He}\ \emph {et~al.}(2012)\citenamefont {He}, \citenamefont {He},\ and\ \citenamefont {Zheng}}]{HeHeZheng}%
  \BibitemOpen
  \bibfield  {author} {\bibinfo {author} {\bibfnamefont {J.-Z.}\ \bibnamefont {He}}, \bibinfo {author} {\bibfnamefont {X.}~\bibnamefont {He}},\ and\ \bibinfo {author} {\bibfnamefont {J.}~\bibnamefont {Zheng}},\ }\bibfield  {title} {\bibinfo {title} {Thermal entangled quantum heat engine working with a three-qubit heisenberg xx model},\ }\href {https://doi.org/10.1007/s10773-012-1084-z} {\bibfield  {journal} {\bibinfo  {journal} {INT. J. THEOR. PHYS.}\ }\textbf {\bibinfo {volume} {51}},\ \bibinfo {pages} {2066} (\bibinfo {year} {2012})}\BibitemShut {NoStop}%
\bibitem [{\citenamefont {Karpat}\ \emph {et~al.}(2020)\citenamefont {Karpat}, \citenamefont {Yal\ifmmode \mbox{\c{c}}\else \c{c}\fi{}\ifmmode \imath \else~\i \fi{}nkaya},\ and\ \citenamefont {\ifmmode~\mbox{\c{C}}\else \c{C}\fi{}akmak}}]{PhysRevA.101.042121}%
  \BibitemOpen
  \bibfield  {author} {\bibinfo {author} {\bibfnamefont {G.}~\bibnamefont {Karpat}}, \bibinfo {author} {\bibfnamefont {i.~d.~I.}\ \bibnamefont {Yal\ifmmode \mbox{\c{c}}\else \c{c}\fi{}\ifmmode \imath \else~\i \fi{}nkaya}},\ and\ \bibinfo {author} {\bibfnamefont {B.}~\bibnamefont {\ifmmode~\mbox{\c{C}}\else \c{C}\fi{}akmak}},\ }\bibfield  {title} {\bibinfo {title} {Quantum synchronization of few-body systems under collective dissipation},\ }\href {https://doi.org/10.1103/PhysRevA.101.042121} {\bibfield  {journal} {\bibinfo  {journal} {Phys. Rev. A}\ }\textbf {\bibinfo {volume} {101}},\ \bibinfo {pages} {042121} (\bibinfo {year} {2020})}\BibitemShut {NoStop}%
\bibitem [{\citenamefont {Peterson}\ \emph {et~al.}(2019)\citenamefont {Peterson}, \citenamefont {Batalhao}, \citenamefont {Herrera}, \citenamefont {Souza}, \citenamefont {Sarthour}, \citenamefont {Oliveira},\ and\ \citenamefont {Serra}}]{WOS:000501493200001}%
  \BibitemOpen
  \bibfield  {author} {\bibinfo {author} {\bibfnamefont {J.~P.~S.}\ \bibnamefont {Peterson}}, \bibinfo {author} {\bibfnamefont {T.~B.}\ \bibnamefont {Batalhao}}, \bibinfo {author} {\bibfnamefont {M.}~\bibnamefont {Herrera}}, \bibinfo {author} {\bibfnamefont {A.~M.}\ \bibnamefont {Souza}}, \bibinfo {author} {\bibfnamefont {R.~S.}\ \bibnamefont {Sarthour}}, \bibinfo {author} {\bibfnamefont {I.~S.}\ \bibnamefont {Oliveira}},\ and\ \bibinfo {author} {\bibfnamefont {R.~M.}\ \bibnamefont {Serra}},\ }\bibfield  {title} {\bibinfo {title} {Experimental characterization of a spin quantum heat engine},\ }\bibfield  {journal} {\bibinfo  {journal} {PHYSICAL REVIEW LETTERS}\ }\textbf {\bibinfo {volume} {123}},\ \href {https://doi.org/10.1103/PhysRevLett.123.240601} {10.1103/PhysRevLett.123.240601} (\bibinfo {year} {2019})\BibitemShut {NoStop}%
\bibitem [{\citenamefont {Hassler}\ \emph {et~al.}(2015)\citenamefont {Hassler}, \citenamefont {Catelani},\ and\ \citenamefont {Bluhm}}]{datos_2005}%
  \BibitemOpen
  \bibfield  {author} {\bibinfo {author} {\bibfnamefont {F.}~\bibnamefont {Hassler}}, \bibinfo {author} {\bibfnamefont {G.}~\bibnamefont {Catelani}},\ and\ \bibinfo {author} {\bibfnamefont {H.}~\bibnamefont {Bluhm}},\ }\bibfield  {title} {\bibinfo {title} {Exchange interaction of two spin qubits mediated by a superconductor},\ }\href {https://doi.org/10.1103/PhysRevB.92.235401} {\bibfield  {journal} {\bibinfo  {journal} {Phys. Rev. B}\ }\textbf {\bibinfo {volume} {92}},\ \bibinfo {pages} {235401} (\bibinfo {year} {2015})}\BibitemShut {NoStop}%
\bibitem [{\citenamefont {Chan}\ \emph {et~al.}(2021)\citenamefont {Chan}, \citenamefont {Sahasrabudhe}, \citenamefont {Huang}, \citenamefont {Wang}, \citenamefont {Yang}, \citenamefont {Veldhorst}, \citenamefont {Hwang}, \citenamefont {Mohiyaddin}, \citenamefont {Hudson}, \citenamefont {Itoh}, \citenamefont {Saraiva}, \citenamefont {Morello}, \citenamefont {Laucht}, \citenamefont {Rahman},\ and\ \citenamefont {Dzurak}}]{datos_2021}%
  \BibitemOpen
  \bibfield  {author} {\bibinfo {author} {\bibfnamefont {K.~W.}\ \bibnamefont {Chan}}, \bibinfo {author} {\bibfnamefont {H.}~\bibnamefont {Sahasrabudhe}}, \bibinfo {author} {\bibfnamefont {W.}~\bibnamefont {Huang}}, \bibinfo {author} {\bibfnamefont {Y.}~\bibnamefont {Wang}}, \bibinfo {author} {\bibfnamefont {H.~C.}\ \bibnamefont {Yang}}, \bibinfo {author} {\bibfnamefont {M.}~\bibnamefont {Veldhorst}}, \bibinfo {author} {\bibfnamefont {J.~C.~C.}\ \bibnamefont {Hwang}}, \bibinfo {author} {\bibfnamefont {F.~A.}\ \bibnamefont {Mohiyaddin}}, \bibinfo {author} {\bibfnamefont {F.~E.}\ \bibnamefont {Hudson}}, \bibinfo {author} {\bibfnamefont {K.~M.}\ \bibnamefont {Itoh}}, \bibinfo {author} {\bibfnamefont {A.}~\bibnamefont {Saraiva}}, \bibinfo {author} {\bibfnamefont {A.}~\bibnamefont {Morello}}, \bibinfo {author} {\bibfnamefont {A.}~\bibnamefont {Laucht}}, \bibinfo {author} {\bibfnamefont {R.}~\bibnamefont {Rahman}},\ and\ \bibinfo {author} {\bibfnamefont {A.~S.}\ \bibnamefont {Dzurak}},\ }\bibfield  {title}
  {\bibinfo {title} {Exchange coupling in a linear chain of three quantum-dot spin qubits in silicon},\ }\href {https://doi.org/10.1021/acs.nanolett.0c04771} {\bibfield  {journal} {\bibinfo  {journal} {Nano Letters}\ }\textbf {\bibinfo {volume} {21}},\ \bibinfo {pages} {1517} (\bibinfo {year} {2021})},\ \bibinfo {note} {pMID: 33481612},\ \Eprint {https://arxiv.org/abs/https://doi.org/10.1021/acs.nanolett.0c04771} {https://doi.org/10.1021/acs.nanolett.0c04771} \BibitemShut {NoStop}%
\bibitem [{\citenamefont {Geyer}\ \emph {et~al.}(2024)\citenamefont {Geyer}, \citenamefont {Het{\'e}nyi}, \citenamefont {Bosco}, \citenamefont {Camenzind}, \citenamefont {Eggli}, \citenamefont {Fuhrer}, \citenamefont {Loss}, \citenamefont {Warburton}, \citenamefont {Zumb{\"u}hl},\ and\ \citenamefont {Kuhlmann}}]{nature_ariel}%
  \BibitemOpen
  \bibfield  {author} {\bibinfo {author} {\bibfnamefont {S.}~\bibnamefont {Geyer}}, \bibinfo {author} {\bibfnamefont {B.}~\bibnamefont {Het{\'e}nyi}}, \bibinfo {author} {\bibfnamefont {S.}~\bibnamefont {Bosco}}, \bibinfo {author} {\bibfnamefont {L.~C.}\ \bibnamefont {Camenzind}}, \bibinfo {author} {\bibfnamefont {R.~S.}\ \bibnamefont {Eggli}}, \bibinfo {author} {\bibfnamefont {A.}~\bibnamefont {Fuhrer}}, \bibinfo {author} {\bibfnamefont {D.}~\bibnamefont {Loss}}, \bibinfo {author} {\bibfnamefont {R.~J.}\ \bibnamefont {Warburton}}, \bibinfo {author} {\bibfnamefont {D.~M.}\ \bibnamefont {Zumb{\"u}hl}},\ and\ \bibinfo {author} {\bibfnamefont {A.~V.}\ \bibnamefont {Kuhlmann}},\ }\bibfield  {title} {\bibinfo {title} {Anisotropic exchange interaction of two hole-spin qubits},\ }\bibfield  {journal} {\bibinfo  {journal} {Nature Physics}\ }\href {https://doi.org/10.1038/s41567-024-02481-5} {10.1038/s41567-024-02481-5} (\bibinfo {year} {2024})\BibitemShut {NoStop}%
\end{thebibliography}
%

\clearpage
\onecolumngrid

\end{document}